\documentclass[onecolumn]{els-mrw} 

\usepackage{amsmath,amssymb,amsfonts,amsthm,makeidx,graphicx}
\usepackage{txfonts}
\usepackage{helvet}
\usepackage{cancel}
\usepackage{braket}
\newcommand{\tr}{\textrm{tr}\,}
\newcommand{\str}{\textrm{str}\,}
\newcommand{\keyword}[1]{\textbf{#1}}
\newcounter{boxcounter}[section]
\renewcommand{\theboxcounter}{\thesection.\arabic{boxcounter}}


\makeatletter
\def\articlecopyright{}
\def\@articletag{}
\makeatother

\begin{document}

\chapter{Quantum chaos and the  holographic principle}\label{chap1}

\author[1]{Alexander Altland}
\author[2]{Julian Sonner}

\address[1]{\orgname{Institute for theoretical physics},  \orgaddress{Zülpicher Str. 77a, 50937 Köln, Germany}}
\address[2]{\orgname{Department of Theoretical Physics}, \orgaddress{University of Geneva, 24 quai Ernest-Ansermet, 1211 Gen\`eve
4, Suisse}}

\maketitle

\begin{glossary}[Keywords]

\end{glossary}

\begin{abstract}[Abstract]
Recent years have witnessed tremendous progress in developing a fine-grained
low-dimensional holographic correspondence, specifically the construction of
quantum mechanical boundary theories as holographic duals of two-dimensional
gravity. In these developments, quantum chaos played a crucial role, both as
source of universality and as a guiding principle for the matching of bulk and
boundary signatures of gravity. In this article we review the construction of
the chaos-assisted low-dimensional holographic correspondence for non-experts.
We open with an introductory discussion of the two main protagonists of the
theory, the SYK model and two-dimensional Jackiw--Teitelboim gravity.
Within this framework we will discuss two independent `bridges' between bulk and
boundary physics, one pertaining to early time chaotic instabilities, the other
to late time quantum chaos up to and including time scales of the order of the
gravitational quantum level spacing. We will demonstrate that the resolution of
these fine-grained quantum scales requires the extension of semiclassical
gravity by elements of string theory. We conclude with an outlook towards higher
dimensional generalizations of the chaotic holographic correspondence. 
\end{abstract}

\section{Introduction}\label{sec:Introduction}

Solving gravity from the foundations of Einstein’s theory to the deep microstate
quantization of black holes remains one of the biggest open problems in physics.
Perhaps the single most important advance in this direction came with the
discovery of the holographic correspondence~\cite{Maldacena1998AdSCFT,Aharony:1999ti,Harlow:2018fse}. It revealed a link between two individually powerful concepts—the
geometric foundations of gravity and quantum field theory—providing a duality
principle and the possibility to validate results obtained in one framework by
comparison to the other. The classical example showcasing the potency of this
connection is Maldacena’s duality between $\mathcal N=4$ super Yang–Mills theory
and gravity in five-dimensional Anti–de Sitter
space~\cite{Maldacena1998AdSCFT,Gubser1998GKPW,Witten1998AdSHolography}.

This paper, for the most part, reviews the more recent construction of a much simpler holographic
correspondence, namely that between strongly interacting one-dimensional
boundary theories (essentially quantum mechanics) and two-dimensional gravity.
This progress arose from the confluence of several ideas, notably the theory of
strongly interacting fermion systems, random matrix theory, two-dimensional
gravity, and topological string field theory. The holographic principle links
these different fields of physics through a single organizing concept: quantum
chaos. The appearance of chaos in the present context should not be surprising.
It reflects the maximally entropic, or chaotic, nature of black
holes~\cite{Natsuume2015AdSCFT,Susskind2016ComputationalComplexity,
Wald1984,HawkingEllis1973, SusskindLindesay2005}.
A distinctive aspect of the two-dimensional holographic correspondence, and one
that will be central in this article, is how quantum chaos turns out to be a
source of universality, revealing connections that might otherwise have remained
opaque.

More concretely, the two-dimensional holographic correspondence revolves around
three theories: the SYK model of $N$ randomly interacting Majorana fermions as a
boundary theory~\cite{KitaevVideo}, the Jackiw–Teitelboim (JT) gravitational path
integral~\cite{Teitelboim1983,Jackiw1984}, and Kodaira–Spencer (KS) field theory
as its string-theoretical completion~\cite{BershadskyCecottiOoguriVafa1994BCOV,
PostvanDerHeijdenVerlinde2022UniverseFieldTheory}. The connection between the
SYK boundary and the JT/KS bulk is established by two independent “bridges.” The
first describes early-time chaotic instabilities at scales $\sim N$, manifesting
themselves in phenomena such as operator scrambling or quantum butterfly
effects~\cite{Maldacena2016,MaldacenaStanfordYang2016NearlyAdS2,
EngelsoyMertensVerlinde2016Backreaction, Jensen2016ChaosAdS2,
PolchinskiRosenhaus2016SpectrumSYK, AlmheiriPolchinski2015AdS2Backreaction}. The
second addresses late time scales, $\sim \exp(N)$, where  chaotic correlations
of
spectra~\cite{CotlerEtAl2017BlackHolesRandomMatrices,SaadShenkerStanford2019JTMatrixIntegral,
StanfordWitten2020JTEnsemblesRMT, Witten2020MatrixModelsDeformationsJT} or even
structures with single level
resolution~\cite{AltlandSonner2021LateTimeHolographicChaos,
AltlandPostSonnerVanDerHeijdenVerlinde2023QuantumChaos2DGravity} take center stage.

\noindent \textit{Less is different:} The anchorage of the bulk–boundary
correspondence in two well-understood theories gives it a high level of
concreteness and makes it the most fine-grained formulation of a holographic
principle known to date. However, to put its development into the context of
holography at large, it is important to realize two distinguishing features of
the low-dimensional setting: first, two-dimensional gravity is conceptually simpler
than its higher dimensional counterparts. Bearing similarity with
two-dimensional electrodynamics, it lacks propagating degrees of freedom
(Einstein equations being of second order their solutions are essentially fixed
by boundary data), bringing topology to the forefront of principles determining
its solutions~\cite{GrumillerKummerVassilevich2002DilatonGravity2D,
GinspargMoore1993Lectures2DGravity}.

Second, the concrete two-dimensional framework discussed here departs from the
orthodox mindset of holography in that it is statistical in nature: it relates
ensemble averages of strongly fluctuating boundary observables to smooth bulk
structures. By contrast, the original holographic principle calls for an exact
equivalence between a microscopically defined boundary Hamiltonian and a bulk
dual. To date, the conceptual status of ensemble averaging remains under active
investigation~\cite{SaadShenkerStanford2019JTMatrixIntegral,StanfordWitten2020JTEnsemblesRMT,
CotlerJensen2022PrecisionTestAveraging, CollierMaloney2022NarainEnsemble}. For
example, one interpretation, to be reviewed later in this paper, rationalizes
the ensemble by considering low-dimensional holography as an effective theory,
obtained by integration over the many degrees of freedom (`ensemble') of
higher-dimensional parent
theories~\cite{PostvanDerHeijdenVerlinde2022UniverseFieldTheory,
AltlandPostSonnerVanDerHeijdenVerlinde2023QuantumChaos2DGravity}. Irrespective
of its origin, the presence of a statistical principle actually has its
own merits: ensemble averaging isolates universal signatures of quantum chaos
after  sample-to-sample information has been averaged out, while retaining
access to fine-grained system information through higher statistical moments. As
a case in point, spectral correlation functions probe the two-dimensional
gravitational spectrum down to  individual black hole microstates, i.e. the
highest possible  level of resolution in this context. At this point, attempts
to lift the insights gained in two to higher dimensions are
underway~\cite{Cotler:2020ugk,Chandra:2022bqq,Belin:2023efa}, but we are only at
the beginning of these developments. 

In this paper, we review the 2D
holographic correspondence in a manner hopefully accessible to both, high- and
low-energy physicists. As always with texts addressing an interdisciplinary
readership, some compromising is involved. For example, we have included some
background material on the two-dimensional AdS black hole or basics of the
statistical theory of quantum chaos, which may be skipped by readers of the
respective camps. In formulas, we often write $X\sim (\dots)$, omitting
numerical factors. The latter can all be found in the cited literature, and we
felt that a  notation emphasizing parametric  connections over
numerical accuracy should have priority in this review.  
Finally, there are
several directions, which we cannot adequately cover in this short review. Specifically, we
will not attempt to review the periodic-orbit/trace-formula approach to
JT/matrix-theory
physics~\cite{Haneder:2024SchwarzianDensity,GarciaGarcia:2019SelbergJT} and the
growing literature on coupling matter to JT
gravity~\cite{Jafferis:2022JTMatter,Blommaert:2019ClocksRods,Moitra:2019ConformalMatterJT};
either of these fascinating topics would require a substantial separate discussion. For a review focused on JT gravity, we refer to~\cite{Mertens:2022irh}. 

Before delving in, let us briefly outline the tale of two-dimensional
holography for the benefit of those who have not heard of it before: 2D gravity
is concerned with the geometry of surfaces. These surfaces can have holes (think
of a torus), or boundaries (a cylinder), or both. While this setting leaves room
for high levels of geometric complexity, the restriction to AdS geometries
imposes the condition of constant negative curvature. Two-dimensional geometries constrained in
this way are still rich, but at the same time simple enough to be quantiatively
describable. In 2D holography, surface boundaries (which have the topology of
circles) are loaded with quantum systems, the boundary coordinate playing the
role of time. These spatially zero-dimensional systems are quantum chaotic, and
the prevalent reading of the situation is that the surface structures connecting
them describe quantum chaotic \textit{correlations} between different replicas
of boundary systems in an ensemble averaged sense. In other words, the surface
geometries encode information otherwise contained in the statistical
correlations of a random matrix ensemble. The common ground on which the two
different perspectives meet is topology. Both random matrix correlations and the
surface structures connecting boundaries can be organized in hierarchies of
ascending topological complexity, and a breakthrough result of the field is the
demonstration that the two approaches quantitatively agree. In this way, chaotic
complexity has been geometrized to infinite order in perturbation theory. The
first five Sections of this review these developments. The final chapter of the
story is the non-perturbative completion of the theory towards a level of
precision where statistical correlations between individual gravitational
quantum states are resolved. We will demonstrate that this  can be achieved
by understanding the bulk surface structures in terms of the  worldsheets of topological
strings. This extension naturally connects to non-perturbative approaches to
quantum chaos via nonlinear $\sigma$-models, completing the
description of gravitational quantum chaotic correlations. 

We begin the detailed narrative with  reviews of its main protagonists, the SYK
model, Section~\ref{sec:syk_model} and JT gravity, Section~\ref{sec:ClassicalJT}.
In Section~\ref{sec:MatrixTheory} we discuss matrix theory from the two
perspectives required in the context of the holographic correspondence: matrices
as proxies of topological geometric structures, and as models for chaotic
correlations. This will form the basis for a comparison between the 
expansion of the JT path integral in gravitational geometries of ascending
complexity and the perturbative $1/N$-expansion of a matrix ensemble in
Section~\ref{sec:TopoJT}. Finally, in
Section~\ref{sec:NonPerturbativeJT}, we will demonstrate how a non-perturbative
completion of the JT path integral is achieved by the inclusion of elements of
string theory. We will discuss how this extension addresses two problems at once,
the statistical ensemble nature of the theory, and the description of the
gravitational spectrum down to the level of individual microstates. We conclude
with an outlook on the extension of the theory to higher dimensions in Section
\ref{sec:Discussion}. 

\section{SYK Model}
\label{sec:syk_model}

The SYK model was proposed in 2015~\cite{KitaevVideo,Maldacena2016} as
holographic boundary dual of two-dimensional gravity. As such it had to tick a
number of boxes expected to be fulfilled by such a model: 1) quantum chaotic
behavior leading to maximal entropy across all relevant time scales, 2) absence of
spatial extension, making it a $0+1$-dimensional system in zero space and one
time-dimension, and conformal symmetry in this time direction to stay close to the
`AdS/CFT' paradigm, which posits conformal boundary theories as partners of AdS
spaces. 

Before discussing the model in the context of these criteria, we note that since
its introduction it has become a
paradigm in the field of  many-body physics at large, with applications in condensed matter
physics, chaos, and gravity. For a number of extensive reviews, we refer to  Refs.~\cite{Rosenhaus2019SYKIntro,
Trunin2020PedagogicalSYKJT,
Sachdev2024QuantumStatMechSYK}.

\subsection{Model definition}

Kitaev  satisfied all three conditions formulated above in terms of a
deceptively simple  model definition involving  $N\gg 1$ Majorana fermions
$\eta_a$, i.e. real fermions with commutation relations 
$[\eta_a,\eta_b]_+=2 \delta_{ab}$, governed by the Hamiltonian
\begin{align}
    \label{eq:SYKHamiltonian}
    \boxed{H=\frac{1}{4!} \sum_{a,b,c,d}^NJ_{abcd} \,\eta_a \eta_b \eta_c \eta_d}
\end{align} 
with Gaussian distributed all-to-all exchange constants $\langle |J_{abcd}|^2
\rangle= 6\frac{J^2}{N^3} $. Note the absence of a
quadratic term, making this a model of `infinitely strong' particle
interactions.  

\begin{BoxTypeA}[box:syk-models]{SYK models}

The SYK model as defined by Kitaev stands in the tradition of an older family
of model systems introduced in the early 1970s to describe the statistical
properties of heavy nuclei~\cite{French1971,Bohigas1971a}. As an alternative to
the, then popular, modelling of  the nuclear
Hamiltonian as a random matrix, this more refined approach started from
the two-body interaction $H=\sum_{\alpha,\dots,\delta=1}^N
J_{\alpha,\dots,\delta} c^\dagger_\alpha c^\dagger_\beta c_\gamma c_\delta$ of
$N$ `hadrons', with $\mathcal{O}(N^4)$ randomly distributed coupling constants
$\{J_{\alpha,\dots,\delta}\}$. In this way,  these models defined ensembles 'embedded' into a
$\sim 2^N$-dimensional Hilbert  
space by a comparatively small number of statistically independent parameters.
However, at the time, the statistical correlations present in the embedded
ensembles were considered to be too complex for analytical solutions. Apart from
the numerical observation of an approximately Gaussian distributed many-body
density of states, their  physics  remained poorly
understood. 

In 1993 Sachdev and Ye~\cite{sachdevGaplessSpinfluidGround1993} rediscovered this
model class --- then in the context of random magnets --- and made breakthrough
contributions to its analytical solution. Focusing on physics at early time
scales $t\sim \mathcal{O}(N)$, they introduced a
Luttinger-Ward~\cite{luttingerGroundStateEnergyManyFermion1960} type functional
integral representation, nowadays known as $G \Sigma$-functionals. Within this
framework, they described manifestations of `maximal many-body chaos', outside
the universality class of the Fermi liquid. Subsequent work~\cite{ParcolletGeorges1999,Sachdev2010} realized the
similarity of the Majorana correlation functions in the non-Fermi liquid regime
to those of conformally invariant systems. 

This insight was brought to fruition by  Kitaev (the `K' in SYK-models) within
the framework of a somewhat simplified model of Majorana, instead of complex
fermions. (The first written account of Kitaev's
introduction of the model in a talk~\cite{KitaevVideo} is
Ref.~\cite{Maldacena2016}.) He realized that the embedded ensembles possess a vast
symmetry under continuous reparameterizations of time. The  crucial impact of
this symmetry on the model's physical behavior will be discussed in the
remainder of this section.      

Finally, a number of generalizations of the original SYK model have been
introduced, and discussed in the holographic context. They include the SYK$_p$
models, defined in terms of Majorana $p$-body interactions as leading
contributions~\cite{Maldacena2016,BerkoozIsachenkovNarovlanskyTorrents2019DoubleScaledSYK},
and tensor models~\cite{Witten2016,KlebanovTarnopolsky2016} which are clean systems, defined to mimic the tensorial
contractions of Majorana fermion operators emerging in the random model after
ensemble averaging.  
\end{BoxTypeA}

Despite its innocent appearance, the SYK Hamiltonian
Eq.~\eqref{eq:SYKHamiltonian} displays extremely rich physics, which, 
uncharacteristically for a strongly interacting system,   is
largely accessible to analytical methods. Specifically, there exist three major
analytical approaches, tailored to different aspects of the system, as
summarized in Table~\ref{tab:1}. While there is no one--does--it--all theory,
the combination of these three provides a rather complete picture of SYK
physics, and it informs the construction of its holographic dual. In the
following three subsections, we briefly review these three approaches in turn.

\begin{table}[h]
    \centering
    \begin{tabular}{|l|l|}
        \hline
        \textbf{$G\Sigma$-theory} & Early time quantum chaos, conformal
        invariance. \\ 
        \textbf{Chord Diagrams} & Collective spectral properties\\
        \textbf{Nonlinear Sigma Model} & Late time quantum chaos, random matrix correlations \\
        \hline
    \end{tabular}
    \vspace{0.5cm}
    \caption{The three principal approaches to SYK physics, and their
    application fields\label{tab:1}. }
\end{table}

\subsection{$G\Sigma$-theory: Emergent conformal symmetry and early time quantum chaos}
Various  physical properties of the SYK model  can be deduced by straightforward scaling analysis:
imagine the model represented in the language of an imaginary
time path integral with symbolic action (we here omit all indices to keep the notation slim) $S\sim \int d\tau
\left( \eta d_\tau \eta - J \eta^4 \right)$, where the first term
implements the canonical commutation relations of Majorana fermions. Assuming
that the interaction constants are large compared to the field fluctuation rates at
relevant time scales, $J\gg d_\tau$, we temporarily forget about the first
term to note that the remainder of the action possesses a vast symmetry:
reparameterizing the time axis as $\tau\mapsto f(\tau)$ in terms of a
diffeomorphism (one-to-one, smooth) $f$, and
redefining the Majorana fields as $\eta(\tau)=\eta'(\tau) f^{\prime 1/4}(\tau)$,
with $f'=d_\tau f$ leaves it invariant. In other words, the system possesses
an approximate symmetry under the infinite dimensional group $\mathrm{Diff}(S^1)$ of diffeomorphisms  of
the the imaginary time circle onto itself.\footnote{Within the finite temperature framework, the
circumference of that circle is set by inverse temperature, $\beta$. However, as
we won't discuss  finite temperature
effects~\cite{bagretsPowerlawOutTime2017a} here, we temporarily focus on 
the limit $\beta\to \infty$.} Note that this symmetry is explicitly broken
upon including the time derivative. Similarly, it would be broken by quadratic
contributions $\sim \eta^2$ to the Hamiltonian. 

\begin{BoxTypeA}[box:ncft]{`Nearly' conformal field theory, NCFT$_1$}

Consider the  SYK symmetry group  $\mathrm{Diff}(S^1)$.  As  a continuous group, it is generated by a
`Lie-algebra', i.e. diffeomorphisms asymptotically close to the identity. For
$f(\tau)=\tau+\epsilon(\tau)$, with small $\epsilon(\tau)\equiv \sum_{l}
\epsilon_l \tau^l$, the action of diffeomorphisms on a space of smooth functions is
represented as $F(\tau)\to F(\tau + \epsilon(\tau))\approx F(\tau)+ \sum_l
\epsilon_l \tau^l d_\tau F(\tau)$, demonstrating that  the operators $L_l \equiv -
\tau^{l+1} d_\tau$ play the role of  group generators. Their commutation relations
$[L_m,L_n]=(m-n)L_{n+m}$, identify them as elements of the \textbf{Witt algebra}. The appearance of the
Witt algebra in the present context contains a number of important messages.
First, it is closely related to the algebra of local conformal transformations in
one-dimension higher, and upon extension the Virasoro algebra,\footnote{To be
precise, a Virasoro algebra with central charge is a unique central extension of
the Witt algebra.} i.e. the symmetry
algebra of two-dimensional conformal field theory. This proximity to $2d$ CFTs
motivated Maldacena to term the effective low energy theory of the SYK model
discusse below a
NCFT$_1$, a \textit{nearly conformal field theory in one dimension}. 

Second, the infinite dimensional Witt algebra contains a miniature
three-dimensional sub-algebra spanned by $\{L_{-1},L_0,L_1\}=\{-d_\tau, -\tau
d_\tau,-\tau^2d_\tau\}$. It is straightforward to check that these generators
commute among themselves, and that their commutation relations are those of
$\mathrm{SL}(2,\Bbb{R})$ aka the three-dimensional group of invertible
unit-determinant matrices $ \left(\begin{smallmatrix} a&b\\ c &d
\end{smallmatrix}\right) $, $ad-bc=1$. In the present context, this group is
represented as $\tau\mapsto (a \tau +b)(c\tau +d)^{-1}$, which are the global
conformal transformations of one dimensional space. As we will see, this
subgroup plays an important role in the theory of our system. 
\end{BoxTypeA}

There is one more thing to be learned from the symbolic representation above: the dimensionlessness of the action, i.e. its  invariance under a local
rescaling $\tau\to b \tau$, requires the fermion fields to have dimension
$[\eta]=\tau^{-1/4}$. This feature puts the SYK model outside the universality
class of the Fermi liquid --- the standard theory of interacting fermions ---
where a dominant quadratic action requires $[\eta]=\tau^{-1/2}$. For example, for
short time separations, $|\tau-\tau'|\lesssim N/J $ SYK correlation functions $\langle
\eta(\tau)\eta(\tau') \rangle\sim |\tau-\tau'|^{-1/2}$, reflecting the dimension $[\eta
\eta]=\tau^{-1/2}$, and different from the textbook $|\tau-\tau'|^{-1}$ of Fermi liquid
quasiparticles. 

\subsubsection{$G\Sigma$-theory}

The above observations are generic in that they apply to any fermion model
subject to asymptotically strong interactions. In order to arrive at a theory
specifically for the SYK universality class, we need chaos, seeded by the randomness
of the interaction constants, enter the stage: 
considering the interaction in its actual form, $\int dt\, H=\frac{1}{4!}\sum J_{abcd}\int
dt\, \eta_a(\tau) \eta_b(\tau) \eta_c(\tau) \eta_d(\tau) $, we begin by
averaging  the imaginary time evolution
$\exp(-\int d\tau\, H)$ over the 
interaction constants. With the Gaussian identity $\langle \exp(X)
\rangle=\exp(\frac{1}{2}\langle X^2 \rangle )$, we obtain the exponent $-\frac{J^2}{8 N^3} \sum \int
d\tau d\tau'\, \eta_a(\tau)\eta_a(\tau') \eta_b(\tau)\eta_b(\tau') \eta_c(\tau)\eta_c(\tau')
\eta_d(\tau)\eta_d(\tau')  $. Nonlocal in time, and of eighth order in the fermion
fields, this  may look like an uninviting
expression. However, it has one weak spot which helps us to make progress
towards a manageable theory: the action solely depends on fermion orbital
`singlets'
\begin{align}
    \label{eq:GMajoranaSinglets}
    G(\tau,\tau')\equiv - \frac{1}{N} \sum_{a} \eta_a(\tau)\eta_a(\tau'). 
\end{align} 
Adopting the
rationale of the Luttinger-Ward functional~\cite{luttingerGroundStateEnergyManyFermion1960}, Sachdev and Ye~\cite{sachdevGaplessSpinfluidGround1993} introduced $G(\tau,\tau')$ as
new time-bilocal integration variables, via a Lagrange multiplier field
$\Sigma(\tau,\tau')$ enforcing the locking Eq.~\eqref{eq:GMajoranaSinglets}. This led
them to a functional integral representation of the SYK model in terms of the two
new effective integration fields, the $G \Sigma$-functional, with its action
\begin{align}
    \label{eq:GSigmaAction}
\boxed{S[G,\Sigma]=-\frac{N}{2}\left(\tr\ln(d_\tau+\Sigma)+ \int dt dt' \,
\left( \frac{J^2}{4} (G(\tau,\tau'))^4 
 +G(\tau,\tau')\Sigma(\tau',\tau)\right)\right) }
\end{align} 
Referring for the derivation of this action to the original references, we note its intuitive
structure: the quartic $G(\tau,\tau')^4$
represents the fermion interaction in the language of $G$-variable and the
linear $G\Sigma$-coupling --- to be read as the matrix product
$\tr(G\Sigma)$ in the bilocal time indices, hence the swapped arguments in
$\Sigma(\tau',\tau)$ --- implements the action of the Lagrange multiplier
$\Sigma$. Finally, the leading tr ln is a remnant of the
integration over fermion fields, which became Gaussian after the introduction of
the $G$-replacement. Here, the field $\Sigma$ enters like a `self
energy' in a theory without quadratic Hamiltonian, motivating the denotation
$\Sigma$-like-self-energy. 

\begin{BoxTypeA}[box:replicas]{Replicas}

The application of functional integral methods to the computation of observables
$ \langle \mathcal{O} \rangle $ averaged over realizations of static randomness
--- such as that present in our random coupling constants --- requires the
introduction of replicas, or alternative methods, to get rid of the normalizing
partition functions in representations $\langle \mathcal{O} \rangle=\left\langle
Z^{-1}(\dots) \right\rangle $. As a consequence, the auxiliary variables inherit
a replica index structure $G^{ab}(\tau,\tau')$, $a,b = 1, \dots, R$. In
applications of the $G \Sigma$-formalism for early time scales $t\sim N$, this
complication is generally swept under the rug. So far, it has not been possible
to extend this framework to one including fluctuations between the different
replica channels, which, however, is essential to the description of late time
quantum chaos. This impasse is one of the essential problems in the construction
of a comprehensive theory describing chaotic correlations of the SYK model
throughout the entire time domain.  
\end{BoxTypeA}

\subsubsection{Stationary phase approach and spontaneous symmetry breaking}\label{sec:SYKspontaneous}

The large factor $N$ upfront the $G\Sigma$-action invites  a stationary phase
analysis.  In the limit of slowly fluctuating field variables $d_t G \ll J\times
G$ already mentioned above, one may vary the action $\delta_\Sigma S = \delta_G
S=0$, under the assumption of a negligible $d_t$ in
Eq.~\eqref{eq:GSigmaAction}. In this limit, the two variational equations assume
the simple form $G(\tau,\tau')=-(1/\Sigma)(\tau,\tau')$ and $\Sigma(\tau,\tau')
= +J^2 (G(\tau,\tau'))^3$, and they afford an intuitive  interpretation: up
to sign conventions, the first states that the Green function
$G\sim(\cancel{d_\tau}-\cancel{H}+\Sigma)^{-1}$ is solely governed by a
self energy which, the second equation, is obtained as a scattering vertex
involving three Green functions. It is straightforward to verify that these two
equations are solved by~\cite{sachdevGaplessSpinfluidGround1993}
\begin{align}
    \label{eq:GSigmaStationaryPhase}
    G(\tau,\tau')=-\frac{b}{J^{1/2}}\frac{\textrm{sgn}(\tau-\tau')}{|\tau-\tau'|^{1/2}},\qquad   
    \Sigma(\tau,\tau')=-b^3 J^{1/2}\frac{\textrm{sgn}(\tau-\tau')}{|\tau-\tau'|^{3/2}},
\end{align}
where $b=(4 \pi)^{-1/4}$ is a numerical constant. Note the consistency
$G(\tau,\tau')\sim \left\langle \eta(\tau)\eta(\tau') \right\rangle\sim
(\tau-\tau')^{-1/2}$ with our earlier scaling argument, demonstrating a
non-Fermi liquid phase at the stationary phase level. In the context of that
discussion, we had also observed that the action in the $d_\tau\to 0$ limit
possesses an invariance under time diffeomorphisms, $\tau\mapsto f(\tau)$, a
symmetry that must carry over to the $G \Sigma$-framework. Indeed, it is
straightforward to verify that Eq.~\eqref{eq:GSigmaAction} is invariant under
the transform 
\begin{align} 
    \label{eq:ReparameterizationTransformation}
   \left. \begin{array}{l}
    G(\tau,\tau')\\
    \Sigma(\tau,\tau')
    \end{array}\right\}\to 
    \left\{\begin{array}{l}
    G([f],\tau,\tau')\equiv
f'(\tau)^{1/4}G(f(\tau),f(\tau'))f'(\tau')^{1/4}\\
\Sigma([f],\tau,\tau')\equiv
f'(\tau)^{3/4}\Sigma(f(\tau),f(\tau'))f'(\tau')^{3/4}
    \end{array}  \right..
\end{align} 
 In other words, alongside the stationary phase solution Eq.~\eqref{eq:GSigmaStationaryPhase}, the
theory is solved by an infinite dimensional manifold of reparameterized
configurations. This is a classical example of spontaneous symmetry breaking:
the symmetry of the action is higher than that of its stationary states. A
closer inspection shows that the stationary phase solutions are invariant under
the subgroup $\mathrm{SL}(2,\Bbb{R})\subset
\mathrm{Diff}(S^1)$ of global conformal transformations $f(\tau)= (a
 \tau +b)/(c \tau +d)$ (see Box~\ref{box:ncft}). As a consequence, we are left with a manifold of non-empty transformations, or
Goldstone modes, 
 $f\in \mathrm{Diff}(S^1)/ \mathrm{SL}(2,\Bbb{R}) $. 

\subsubsection{Schwarzian action}\label{sec.SYK-Schwarzian}
Transformations of the stationary phase solutions under the $f$-modes will be
 weighted by a finite action, once the so-far neglected time derivative is
taken into account. (The latter plays a role analogous to that of an external
field in a magnetic symmetry breaking scenario.) On general grounds, we expect
this  Goldstone mode
 action to be (a) of leading order in time derivatives,
 $\mathcal{O}(N d_\tau 
 /J)$, and (b) invariant under global $\mathrm{SL}(2,\Bbb{R})$ conformal transformations. Referring for
its detailed derivation to Ref.~\cite{bagretsSachdevYeKitaev2016} and a more
heuristic discussion to~\cite{Maldacena2016}, these requirements are uniquely met by the
 \textbf{Schwarzian action}:
\begin{align}
    \label{eq:SchwarzianAction}
   \boxed{S[f]=M \int d \tau \{f(\tau),\tau\}, \qquad \{f(\tau),\tau\} \equiv
    \left( \frac{f''(\tau)}{f'(\tau)} \right)'  - \frac{1}{2}\left(\frac{f''(\tau)}{f'(\tau)}\right)^2}
\end{align} 
where $M\sim N/J$. The appropriate measure, $Df$, invariant under the group
action $Df=D(f\circ g)$ is given by $Df=\prod_\tau df(\tau)(f'(\tau))^{-1}$.~\cite{bagretsSachdevYeKitaev2016}

Eq.~\eqref{eq:SchwarzianAction} determines the effective low-energy action of
the SYK model.  Majorana correlation functions can be computed from it via a
three-step protocol: (i) expressing the fermion bilinears in terms of
$G$-variables, (ii) substituting the generalized stationary phase solution
Eq.~\eqref{eq:ReparameterizationTransformation}, and (iii) averaging over the
reparameterization modes $f$ with the Schwarzian action. In practice, however,
this program is met with some challenges related to the nonlinearity of the
Schwarzian action and that of the measure. 

The freedom of working with different parameterizations of the time
domain holds the key to overcoming these obstacles: a change of variables,
$f'(\tau)=\exp(\phi(\tau))$, has the double-effect of bringing the action to
form $S[\phi]=M \int d \tau \,\phi^{\prime 2}$, and flattening the measure to
$D\phi=\prod_\tau d\phi(\tau)$. In this representation, the computation of the 
SYK partition function reduces to that of a free quantum particle of mass $M$.
The problem gets a little more involved when correlation functions, such as
expectation values of Green functions or their moments are considered: in
$\phi$-language, the algebraic denominators emerging when
Eq.~\eqref{eq:GSigmaStationaryPhase} is substituted into
Eq.~\eqref{eq:ReparameterizationTransformation}  assume the form
\begin{align*}
    \frac{1}{(f(\tau)-f(\tau'))^{1/2}}= \left( \int_{\tau'}^{\tau} d s \, e^{\phi(s)} \right)^{-1/2} =\frac{1}{\sqrt{\pi}} \int_0^\infty \frac{d \alpha}{\sqrt{\alpha}}e^{-\alpha \int_{\tau'}^{\tau} d s \, e^{\phi(s)} }.
\end{align*} 
In other words, our problem  is effectively described by the action 
\begin{align}
    \label{eq:LiouvilleAction}
    S[\phi]=\int d\tau \left( M \phi^{\prime 2} + \alpha e^{\phi} \right).
\end{align}
Eq.~\eqref{eq:LiouvilleAction} describes the  quantum mechanics of a particle of
mass $M$ subject to an exponential potential $V(\phi)=\alpha e^{\phi}$, the
so-called \textbf{Liouville quantum mechanics}. 

There are two immediate insights following from this representation: first, the exponential potential acts as an 
impenetrable wall at large positive $\phi$, i.e. Liouville quantum mechanics
essentially describes a free quantum particle confined to the negative
half-plane $\phi<0$, a very simple problem. Second, for characteristic time
differences $\Delta \tau \equiv |\tau-\tau'|$, the action cost of the lowest
lying $\phi$-fluctuations scales as $\sim M \Delta \tau^{-1}$, suggesting that
reparameterization-$\phi$ fluctuations become strong for $\Delta \tau
\gtrsim M \sim N/J$. This timescale marks the crossover from early time
SYK physics, governed by the stationary phase solution
Eq.~\eqref{eq:GSigmaStationaryPhase}, to a later regime dominated by strong
quantum fluctuations of the reparameterization modes.

\subsubsection{Observables}

From this point onward, the computation of SYK observables is conceptually
straightforward, but technically too involved to be reviewed in detail here.
Instead, we just summarize a number of key results that have been obtained by
integrating over the reparameterization modes:

\paragraph{Averaged Green function:} The ensemble averaged Green function, while
not an observable in itself, is an essential building block of SYK physics. We
have already seen that at short timescales it displays non-Fermi liquid scaling
$\langle G(\tau,\tau') \rangle \sim |\tau-\tau'|^{-1/2}$. At larger time
separations $\tau-\tau'\gtrsim N/J$, reparameterization fluctuations set in,
modifying the power law to $|\tau-\tau'|^{-3/2}$, a qualitative change in the
infrared behavior of the theory reflecting the presence of strong quantum
fluctuations. For later reference, we note that this $-3/2$ power law is in line
with a beautiful universality of Liouville quantum mechanics~\cite{KoganMudryTsvelik1996,SheltonTsvelik1998}: for this system,
correlation functions of \textit{all} operators display the same scaling
behavior, $\langle \mathcal{O}(\tau) \mathcal{O}(0) \rangle \sim |\tau|^{-3/2}$.

\paragraph{Out-of-time-order correlation functions} Out of time correlation
functions, such as 
\begin{align}
    \label{eq:OTOCDefinition}
    F(t)=\frac{1}{N^2 Z} 
     \sum_{ab}\left\langle \tr\left(e^{-\frac{\beta}{4} H}\chi_a(0)e^{-\frac{\beta}{4} 
     H}\chi_b(t)e^{-\frac{\beta}{4} H}\chi_a(0)e^{-\frac{\beta}{4} H}\chi_b(t)\right)
\right\rangle ,
\end{align} 
with $Z=\left\langle \tr(\exp(-\beta H)) \right\rangle $ probe the physics of early time information
scrambling and chaos~\cite{Maldacena2016,MaldacenaStanfordYang2016NearlyAdS2,KitaevSuh2018SoftMode}. Here, the quasi-thermal factors $\exp(-\beta H/4)$, evenly
distributed between Majoranas, serve as `filters' which in the limit of low
temperatures $T=\beta^{-1}$ limit the dynamical probe to correlations near the
ground state.  Noting that this correlation function contains pairs of Green
functions as  building
blocks, it becomes amenable to  the protocol outlined above. As a result
of this analysis one finds:
\begin{itemize}
   \item A regime of  initial exponential dephasing $F(t)\sim 1- \frac{1}{64 \pi}
   \exp(2 \pi T(t-t_\textrm{E}))$. This is the famous result of `\keyword{maximal chaos}'~\cite{MaldacenaShenkerStanford2016BoundChaos}: no
   matter how low temperature, the SYK model scrambles information exponentially
   with a Lyapunov exponent $\lambda=2 \pi T$. Note that the state space
   accessible to a system kept at
   temperature $T$ is limited by that temperature, i.e. the rate $\lambda$ defines the
   shortest possible timescale: maximal chaos down to arbitrarily small energy
   scales above the ground state. This scrambling phase is 
   completed at the so-called \keyword{Ehrenfest time} $t\sim t_\textrm{E}\equiv
   \ln(MT)/2\pi T$. For larger times, a large $N$ `semiclassical' stationary
   phase analysis  predicts exponential decay,
   $F(t)\sim \exp(-\pi T (t-t_\textrm{E}))$,~\cite{bagretsPowerlawOutTime2017a},
    consistent with the expected behavior of gravitational shock waves.~\cite{MaldacenaStanfordYang2016NearlyAdS2}
   \item However, this is not the end of the story: at times $t\gtrsim M \propto N \gg t_\mathrm{E}\sim \ln
   N $ quantum fluctuations of the reparameterization modes become strong. In this regime, the OTOC crosses over to a power law decay,
   $F(t)\sim t^{-6}$.~\cite{bagretsPowerlawOutTime2017a} In view of the abovementioned Liouville universality, this
   finding is not too surprising: the fully connected correlation function
   $F(t)$ involves four quantum amplitudes between Majorana operators at times
   $0$ and $t$, respectively. Each contributing a universal power law, we end up
   with $t^{-3/2\times 4}=t^{-6}$ decay. Also, note that the timescale $t\sim
   M$ marking the onset of strong quantum fluctuations is inversely proportional to the
   `effective $\hbar $' of the theory, justifying the terminology,
   $t_\mathrm{E}\sim -\ln(\hbar)\sim \ln M $ for the Ehrenfest time. These power
   law long time tails are consistent with, indeed required, by the spontaneous
   breaking of a continuous symmetry. However, on the basis of the same argument
   one should expect that in our one-dimensional system these fluctuations
   ultimately become so strong as to `restore' the spontaneously broken
   reparameterization symmetry. If and how this happens remains an unsolved question.
\end{itemize}

\noindent We conclude this section with a comment on one more conceptually
interesting aspect of Liouville quantum mechanics. Describing a single quantum mechanical
 degree of freedom, it does not show any signatures of
chaos. At the same time, it features as the effective theory of a
maximally chaotic many-body system. In other words,  Liouville quantum mechanics \textit{describes chaos} without being chaotic itself.
Theories of this type are notorious in low dimensional holography. In order to
avoid repetition of the clumsy attribute
'describes-chaos-without-being-chaotic-itself', we will refer to them  using the tags defined in
Table \ref{tab:2}. An important question currently under investigation is
whether low-dimensional holography is entirely formulated in terms of
CC-theories, or whether individual realizations of CH-theories feature, e.g. as
`ultraviolet closures'. We will return to this question in the final section of
this review.  

\begin{table}[h]
    \centering
    \begin{tabular}{|l|l|}
        \hline
        \textbf{CH}-theory & A quantum-\textbf{CH}aotic theory (e.g. single realization of
        the SYK Hamiltonian).\\
        \textbf{CC}-theory & A theory describing \textbf{C}haotic
        \textbf{C}orrelations. (e.g. Liouville quantum mechanics).\\
        \hline
    \end{tabular}
    \vspace{0.5cm}
    \caption{Acronyms for theories describing chaos used in this paper.\label{tab:2}}
\end{table} 

\subsection{Chord diagrams: The many body spectral density of the SYK model }
\label{sec:ChordDiagrams}
Previously we have seen that the SYK model displays quantum chaos down to
 energy scales arbitrarily close to the ground state. The holographic
correspondence unfolds precisely in this regime, and it makes extensive reference to the
near ground state \textit{spectral} properties of the SYK Hamiltonian. Of particular
interest in this context are the `entropic properties' of the model at low
temperatures, essentially the number of levels in the immediate vicinity of the
ground state. If the  SYK Hamiltonian were
describable in terms of a $D\propto \exp(N)$ dimensional Gaussian distributed random matrix, then we should
expect a `Wigner semicircular' near edge spectral density, $\rho\sim
D\sqrt\epsilon$, and the large prefactor would translate into a low temperature
entropy $S\sim D\sim N$. The question is whether this expectation, which is incompatible
 with the vanishing of the zero temperature entropy shown by generic
many body systems, is an artifact of the random matrix approximation, or whether
the SYK Hamiltonian indeed displays a near edge spectral density of this type.

\begin{figure}[t]
\centering
\includegraphics[width=1\textwidth]{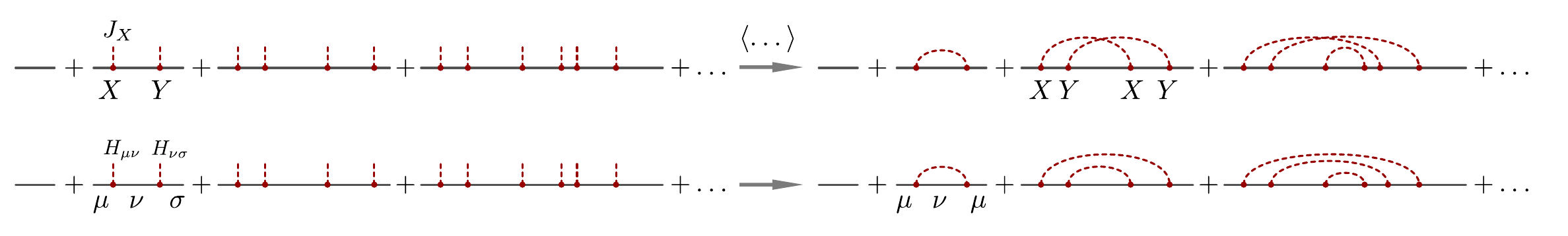}
\caption{Top: Chord diagrams describing the configurational 
average of a Majorana two-point function scattering off Fock space operators $X=
\eta_a \eta_b \eta_c \eta_d$ with random couplings $J_{abcd}$. Averaging
enforces the pairwise occurrence of identical $X$-operators where the
bookkeeping is done via chord connectors.  Bottom: the analogous diagram for the
scattering off a Gaussian distributed random matrix Hamiltonian $\left\langle
H_{\mu_\nu}H_{\nu'\mu'} \right\rangle \propto \frac{1}{D} \delta_{\mu
\mu'}\delta_{\nu \nu} $ in the limit of large Hilbert space dimension, $D$.
Here, the condition to obtain a maximal number of free running index summations
enforces a non-crossing rule in the averaged theory. }
\label{fig:ChordDiagrams}
\end{figure}

To answer it, we need a framework for the computation of the SYK
spectral density. It turns out that this task is best addressed in a `first
quantized' setting, interpreting the SYK Hamiltonian as a $D\times D$ matrix in
the $(D=2^{N/2})$-dimensional Hilbert space of $N/2$ complex fermions defined
by $N$ Majorana fermions. Within this space, the Hamiltonian is realized as
$H=\sum_X J_X X^{-1}$, with  operators $X\equiv \chi_a
\chi_b \chi_c \chi_d $ and random couplings
$J_X\equiv J_{abcd}$. Now consider  the  density of states represented as 
\begin{align*}
    \rho(E)=-
\frac{1}{\pi}\mathrm{Im} \left\langle \tr \left( \frac{1}{E^+-H}  \right)
\right\rangle = 
\frac{1}{\pi}\mathrm{Re} \int_0^\infty dt \,e^{i E^+ t} \left\langle \tr \left(
e^{-iH t}  \right)\right\rangle = \frac{1}{\pi}\mathrm{Re} 
\int_0^\infty dt \,e^{i E^+ t}\sum_{l}  \frac{(-it)^{2l}}{(2l)!} 
\left\langle \tr 
\left( \sum_X J_X X \right)^{2l}   \right\rangle,
\end{align*} 
i.e. a formal real time series as indicated in the upper half of
Fig.~\ref{fig:ChordDiagrams}, where the dummy horizontal legs may be interpreted
as propagation under some `null Hamiltonian' between scattering events (or just be
forgotten about). The Gaussian distribution of the couplings $J_X$
implies that only even powers, $2l$, feature in the expansion, and averaging
over it enforces pairwise equal $X$-operators, weighted with the variance $6J^2/N^3$. 
The different $(2l-1)!!$ possible pairings can be distinguished by connecting
lines (Fig.~\ref{fig:ChordDiagrams}, top right), hence the name \keyword{chord diagrams}.~\cite{BerkoozNarayanRaz2020SYKChordDiagrams} 

To turn this into a controlled expansion of the spectral density in $N\gg 1$, we
observe that in the limit $N\to \infty$ different $X$-operators typically do
not share a common Majorana operator, and hence commute. Also note $X^2=1$ as a consequence of
the Majorana anti-commutation relations. Exploiting this property,
we have $\tr(X \dots Y \dots X \dots Y \dots)\approx \tr(1)=D$. The now free running
summations $\sum_X 1 =\sum_{a<b<c<d}1 =\binom{N}{4}\equiv K$ produce
counting factors and we arrive at 
estimate
\begin{align}
    \label{eq:GaussianDoS}
    \rho(E)\approx  \frac{D}{\pi} \textrm{Re}  \int_0^\infty dt \,e^{i E^+ t} \sum_l \left( -it \right)^{2l} \left( \frac{K 6J^2}{N^3}  \right)^l  \frac{(2l-1)!!}{(2l)!} =
    \frac{D}{2\pi}   \int_{-\infty}^\infty dt \,e^{i E^+ t}e^{ \frac{-(\gamma t)^2}{2} } =  \frac{D}{\gamma\sqrt{2 \pi}}
    \exp \left( - \frac{E^2}{2 \gamma^2}  \right).
\end{align} 
where 
\begin{align}
    \label{eq:SYKBandWidth}
    \gamma=\left(  \frac{6J^2 K}{N^3}  \right)^{1/2}\sim JN^{1/2}
\end{align} 
is a measure of the \keyword{SYK bandwidth}. Its scaling form can be understood
from a simple  heuristic argument: the SYK Hamiltonian contains $\mathcal{O}(N^4)$
    terms of characteristic magnitude $\sim \textrm{rms}(J_X)\sim J/N^{3/2}$,
    the central limit theorem then indicates a bandwidth  $\gamma\sim N^2
    J/N^{3/2}\sim JN^{1/2}$, consistent with~\eqref{eq:SYKBandWidth}. 

The Gaussian envelope function Eq.~\eqref{eq:GaussianDoS} is a hallmark of many
    body quantum chaos. In it reflects the fact that, unlike the  Gaussian
    distributed random matrix Hamiltonian with its semicircular spectral
    density, many-body Hamiltonians do not maximize entropy. The `information'
    carried by the Hamiltonian is contained in the commutator algebra of its
    constituent operators, $X$, which in turn determines the series structure
    leading to the Gaussian spectral density. (For comparison, the expansion of
    a Gaussian distributed maximum entropy Hamiltonian $H=\{H_{\mu \nu}\}$ is
    visualized in the bottom part of Fig.~\ref{fig:ChordDiagrams}. Matrices
    drawn from this ensemble generally are `full', implying that contributions
    to the expansion maxing out the number of free running index summations must
    have non-intersecting chords. The resummation of this series leads to the
    celebrated Wigner semicircle.)

Going beyond the crude approximation $[X,Y]=0$, chord diagram analysis has
turned into an art form producing  powerful results for the near edge spectral
density, and its fluctuations:
\begin{itemize}
    \item A straightforward generalization of the series representation above,
    still neglecting commutators, shows that the spectral density is subject to
    `collective' sample-to-sample fluctuations with variance
    $\mathrm{var}(\rho(E))\sim \frac{\langle \rho(E) \rangle^2}{2K}  $, much
    larger than the almost absent fluctuations of the random matrix spectral
    density.~\cite{JiaVerbaarschot2019SpectralFluctuationsSYK} (Here, the
    attribute `collective' means that we are discussing the spectral density
    binned over many level spacings, different from the `microscopic'
    fluctuations at level spacing scales.) These fluctuations originate in the
    fact that the SYK Hamiltonian contains only $\mathcal{O}(N^4)$ random
    parameters, each affecting exponentially many correlated matrix elements.
    Referring for a detailed discussion  to
    Ref.~\cite{altlandQuantumChaosEdge2024}, we  call this \keyword{sparse
    quantum chaos}, different from the \keyword{dense quantum chaos} realized by
    the Gaussian random matrix Hamiltonian. 
    \item The  constancy of the total number of levels indicates that these
    fluctuations must induce equally strong shifts of the spectral edge, $E_0$,
    i.e. the system's ground
    state energy. Indeed, it turns out that the latter is distributed with
    variance $\textrm{var}(E_0)\sim (JN)^2/2K$.~\cite{BerkoozBruknerNarovlanskyRaz2021MultiTraceSYK}
    \item The inclusion of higher order commutators shows that the average
    spectral density of the SYK Hamiltonian
    is given by~\cite{GarciaGarciaVerbaarschot2016SYKSpectralThermo}
    \begin{align}
        \label{eq:SYKDoS}
        \left\langle \rho(E) \right\rangle = e^{S_0}\left\langle 
            \sinh \left( \sqrt{\frac{E-E_0}{c JN}}  \right) \Theta(E-E_0) \right\rangle_{E_0},
    \end{align} 
    where $c$ is a numerical constant, and $S_0 \sim N$ the ground state
    entropy. Interestingly, the $\sinh$-profile can be alternatively
    obtained~\cite{bagretsPowerlawOutTime2017a, StanfordWitten2017Schwarzian}
    from Liouville theory, representing the SYK partition sum as $Z=\tr\exp(-\beta H)=\int
    dE \,\rho(E)\exp(-\beta E)$, and comparing with the results obtained by
    integrating over reparameterization modes. In this context, $\rho(E)$ is a
    measure for the energy stored in Goldstone mode fluctuations. The
    complementary chord diagram analysis adds the important information of edge
    fluctuations. 
    \item We can intentionally get closer to a dense scenario considering model
    variants with leading order interaction of order $p>4$ in the number of
    Majoranas, the SYK$_p$ model mentioned in Box~\ref{box:syk-models}. As $p$ is
    increased, the number of independent operators $X$ increases as
    $\binom{N}{p}\sim N^p$, and the system approaches the dense limit. Indeed,
    chord diagram analysis shows that in the scaling limit $p^2\sim N$ the
    collective fluctuations are suppressed, and accordingly the spectral edge
    ceases to fluctuate. This limiting model is known as the \keyword{double
    scaled SYK model}.~\cite{BerkoozIsachenkovNarovlanskyTorrents2019DoubleScaledSYK}
\end{itemize}
  
\subsection{Nonlinear $\sigma$-model: Microscale spectral correlations}

The final ingredient we need for the discussion of the holographic
correspondence is a characterization of near-edge spectral correlations at the
resolution 
of the many-body level spacing. The method of choice for this
task is the nonlinear $\sigma$-model~\cite{Altland2023,Efetbook}, combined with elements of  chord diagram
analysis. The nonlinear $\sigma$-model adds to our so-far discussion the
principle of spontaneous breaking of causal symmetry characterizing effectively
irreversible chaotic dynamics.~\cite{Wegner1979} Referring to
Ref.~\cite{altlandLateTimePhysics2021} for its discussion in the context of
holographic duality, the statement is that in regions $E>E_0$ with a
non-vanishing spectral density, any amount of averaging breaks the symmetry
between retarded and advanced Green functions $G^\pm(E)=(E\pm i \delta
-H)^{-1}$, in the sense that $\langle \rho(E) \rangle = - \frac{1}{2\pi} \langle
\tr(G^+(E)-G^{-}(E)) \rangle >0 $. In this reading, the spectral edge $E_0$
marks a \keyword{causal symmetry breaking} phase transition point, with the
average spectral density as order parameter.~\cite{altlandQuantumChaosEdge2024} The concomitant Goldstone modes
then mediate universal spectral correlations at the many-body level spacing
scale.  

The theory describing this phenomenology starts from the Gaussian partition sum
\begin{align}
    \label{eq:GaussianFunctional}
    &Z(\hat \epsilon)=\int d \psi\, \left\langle\exp(-S[H,\psi,])\right\rangle,\cr 
    &\quad S[H,\psi]\equiv  i \bar \psi (\hat \epsilon-H) \psi,
\end{align}
where $H=\{H_{\mu,\nu}\}$ is the SYK Hamiltonian realized as a matrix in the
$D$-dimensional space of Fock space states $\braket{\mu}$ (`first quantized'
representation). The integration variables 
$\psi=\{\psi_\mu^\alpha\}$, $\alpha=(a,s)$ are complex commuting
($\psi^{\textrm{b},s}$) or Grassmann valued ( $\psi^{\textrm{f},s}$ ), with
causal indices $s=\pm 1$, and
$\hat \epsilon=\mathrm{diag}(\epsilon^{\textrm{b},s},\epsilon^{\textrm{f},s}) $
is a diagonal matrix of energy arguments.  In this way it is guaranteed that for
$\epsilon^{\textrm{b},s}=\epsilon^{\textrm{f},s}$ the integral is
unit-normalized by supersymmetry (i.e. the integral over Grassmann/commuting
variables producing determinants/inverse determinants, which cancel out), and
differentiation 
with respect to elements of $\hat \epsilon$
(  $\mathrm{Im}\,\hat \epsilon = i \delta \tau_3$)  at the configuration  $\hat
\epsilon = E + i \delta $ generates traces of the retarded or advanced Green
function. 

In this representation, the causal symmetry is realized as  invariance under
transformations $\psi\to T \psi$, $\bar \psi \to \bar \psi T^{-1}$, where $T$ is
a matrix acting in the causal and boson/fermion space. Averaging over the SYK
Hamiltonian generates terms bilinear in the $X$-operators defined in the
previous section (i.e. terms quartic in $\psi$-variables) which are subsequently
decoupled via a Hubbard-Stratonovich transformation. Referring to
Ref.~\cite{altlandQuantumChaosEdge2024} for details, a  stationary phase
analysis then leads to the aforementioned symmetry breaking: a mean field,
proportional to $\textrm{sgn}(\pm i \delta)\equiv i \delta \tau_3$ reducing the
symmetry transformations down to $T_+ \oplus T_-$, diagonal in causal space.
According to the standard symmetry breaking paradigm --- however, the present
form of symmetry breaking should not be confused with the physically
unrelated short timescale breaking of conformal invariance --- the Goldstone
modes of causal symmetry breaking are realized as matrices $Q=T \tau_3 T^{-1}$.
To derive their  effective action one then considers the case of small energy
differences $\omega \equiv \epsilon^+-\epsilon^-$ and expands to leading order
in this symmetry breaking parameter, where traces involving $X$-operators are
evaluated using chord diagram methods. The end result of this procedure is the
nonlinear $\sigma$-model action~\cite{altlandQuantumChaosEdge2024}
\begin{align}
    \label{eq:ActionCollectiveFluctuations}
    S[Q]= \ln  \left \langle \exp\left( -i \pi \rho(E) \frac{\omega}{2} 
    \, \mathrm{str} (Q \tau_3)\right) \right \rangle_\textrm{coll},
\end{align} 
where the averaging is over the collective fluctuations of the spectral density
$\rho(E)$ discussed in the previous section, and $\mathrm{str}(\dots)$ is the
supertrace, i.e. $\str(A)=\tr(A^\mathrm{bb}-A^\textrm{ff} )$.

The exponential in Eq.~\eqref{eq:ActionCollectiveFluctuations} is 
the supersymmetric matrix action from which the second moment of the
spectral density $\langle \rho(E+\frac{\omega}{2} )\rho(E-\frac{\omega}{2} )
\rangle$ can be obtained by integration over $Q$-fluctuations.~\cite{Efetbook} The result of
this procedure is the standard Wigner-Dyson correlation function of the unitary
ensemble (for more details on this point, see Box~\ref{box:sigma-model}). Fourier
 transformation of this result leads to the `ramp' and `plateau' structure of
the spectral form factor, $K(\tau)=\tau \Theta(1-\tau)+ \Theta(\tau-1) $, where
$\tau$ is Fourier conjugate to $s=\pi \omega \rho(E)$, i.e. dimensionless
energy in units of the spectral density at the center energy.

\begin{figure}[t]
\centering
\includegraphics[width=.5\textwidth]{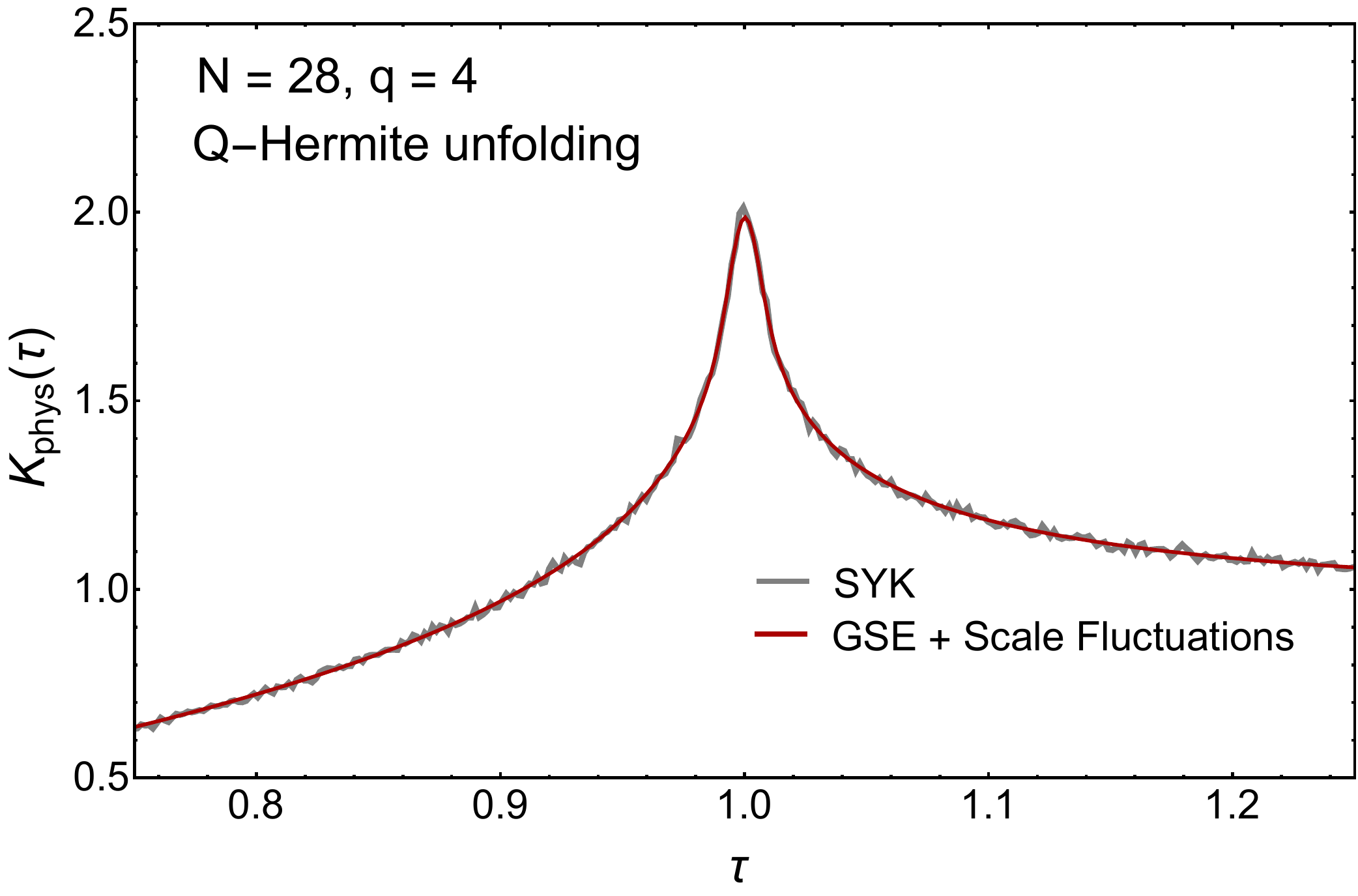}
\caption{The form factor of an $N=28$ SYK model, which is in symmetry class AII, 
the symplectic Wigner-Dyson class. (Depending on the value of $N
\,\textrm{mod}\,8$, the SYK Hamiltonian falls into the unitary, orthogonal or
symplectic class.~\cite{YouLudwigXu2017SYKBoundarySPT})  Away from
$\tau=1$, the spectral form factor  of  
the SYK model, averaged over 10000 realizations, is nearly indistinguishable
from that of the Gaussian Symplectic Ensemble (GSE). However, while the latter
shows a sharp singularity at the Heisenberg time, $\tau=1$, this singularity is
rounded in SYK spectral statistics, here averaged over 10000 realizations. This
result is  in excellent agreement
with the predictions following from Eq.~\eqref{eq:ActionCollectiveFluctuations}.
(Reproduced from Ref.~\cite{altlandQuantumChaosEdge2024}.)}
\label{fig:GSEFormFactor}
\end{figure}

The new element
in our context is that the latter fluctuates. In essence, the average over these
fluctuations translates to an average over $\tau$ over scales $\sim K^{-1/2}\sim
N^{-2}$. Its most visible effect is a rounding of the form
factor over `collective' time scales $\Delta \tau \sim N^{-2}$, replacing the sharp
corners of the ramp and plateau by smooth crossovers, cf.
Fig.~\ref{fig:GSEFormFactor} for an example.  A key
question we will need to address in the following is to what extent these
collective fluctuations are present in the gravitational bulk supposedly dual to
the SYK model.   

\section{JT gravity} \label{sec:ClassicalJT}

In this section we we introduce two-dimensional JT gravity assuming no
prerequisites except familiarity with basic concepts of relativity. The
conections to the previously discussed structure at the boundary will be drawn
in Section~\ref{sec:TopoJT}, after some required background in matrix theory has been
introduced in Section \ref{sec:MatrixTheory}.

\textbf{Jackiw-Teitelboim gravity}, or in short JT gravity, is a two-dimensional theory of gravity governed the action
\begin{equation}\label{eq:JTAction}
    \boxed{S = \frac{1}{16\pi G_N} \int d^{2} x \sqrt{-g}\Phi \left(  R - \Lambda\right) + S_{\rm bd}}
\end{equation}
\begin{figure}[h]
\centering
\includegraphics[width=0.5\textwidth]{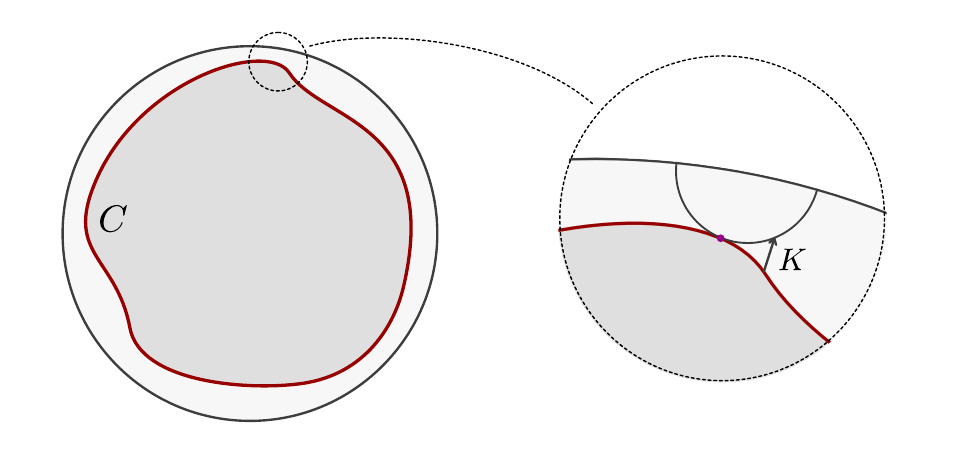}
\caption{A gravitational manifold with boundary  represented as a subset of
the Poincaré disk. Heuristically, the extrinsic curvature of the boundary is
proportional to the rate at which it deviates from the best approximation of a
straight line, a geodesic. The latter assume the form of semicircles which,
reflecting the diverging metric area,  hit the boundary in radial direction. }
\label{fig:JTBoundary}
\end{figure}

Here,  $(-g)$ denotes the (negative) determinant of the spacetime metric
$g_{\mu\nu}$, where $\mu, \nu =0,1$ are Lorentzian spacetime indices, $R$ is the
Ricci scalar of the metric, and $\Lambda$ is the cosmological constant, which we
will choose to be negative for the purposes of this review. The so-called
dilaton field $\Phi$ is a special feature of this two-dimensional model of
gravity, compared to standard Einstein-Hilbert gravity in higher dimensions, and
we will explain its meaning shortly. For the rest of this review, we use units in
which $8\pi G_N = 1$, so that the JT action and its boundary term simplify to
$\frac{1}{2}\int\sqrt{-g}\,\Phi(R-\Lambda)$ and $\int\sqrt{-\gamma}\,\Phi(K-1)$,
respectively. The definition of the theory is completed
by a \textbf{boundary action}, required to make the 
variational problem of \eqref{eq:JTAction}  well posed:
\begin{equation*}  
\boxed{S_{\rm bd} = \int dx \sqrt{-\gamma} \Phi\left(K-1 \right)}
\end{equation*}
In this expression, $\gamma$ is the induced boundary metric,  $K$ its extrinsic
curvature. See Fig.~\ref{fig:JTBoundary} for a heuristic interpretation as well
as \eqref{eq:ExtrinsicCurvature} below for its precise mathematical definition.
The integral extends over the boundary of the two-dimensional manifold, now
parameterized by a one-dimensional coordinate $x$. The second term, akin to a
boundary negative cosmological constant is needed to render correlation
functions finite once extrapolated to the boundary, in a procedure  known as
holographic renormalization (cf, e.g., Ref. \cite{Skenderis:2002wp}.)    

To complete the action of JT gravity, one also has the
freedom to add a term proportional to the Euler characteristic
\begin{equation}\label{eq:EulerCharacter}
   \chi\left( {\cal M} \right) =\frac{1}{4\pi}\left(\int_{\cal M} \sqrt{g}\, R
   + 2\int_{\partial\cal M} \sqrt{\gamma}\, K\right),
\end{equation}
which by the Gauss--Bonnet theorem equals $\chi = 2-2g-n$ for a surface of genus
$g$ with $n$ boundaries. Coupled with a constant $\Phi_0$, the resulting
topological action $S_0\chi$ (with $S_0 = 2\pi\Phi_0$ in our units) will play a
crucial role in our later analysis.

\subsection{Classical solutions and configuration space}\label{sec.ClassicalSols} Before
moving on to quantization, let us establish some features of the classical
theory. The dilaton equation of motion sets curvature, such that the Ricci
scalar $R=\Lambda$. For negative $\Lambda$, this means that the metric is
locally that of AdS$_2$, which in Poincaré coordinates $(\tau,z)$ reads
\begin{equation}
    \label{eq:AdSMetricPoincare}
    ds^2 = \frac{d\tau^2 + dz^2}{z^2}\,,\qquad z\ge 0.
\end{equation}
We work in Euclidean signature, $\tau = it$, and have normalized the curvature
such that $R=-2$. The metric equation of motion then determines the dilaton
field,
$
    \nabla_\mu \nabla_\nu \Phi -g_{\mu\nu} \nabla^2 \Phi + g_{\mu\nu} \Phi =0
$.
For $g_{\mu\nu}$ the constant negative curvature metric above, this has the
general solution
\begin{equation}
    \label{eq:DilatonBulkSolution}
    \Phi = \frac{a + b \tau + c(\tau^2 + z^2)}{z}\,,\qquad a,b,c \in \mathbb{R}.
\end{equation}
All classical solutions are locally AdS$_2$, they merely differ by the
functional form of the dilaton, and the choice of a boundary curve $C$ embedded
in the Poincaré disc (see Fig.~\ref{fig:JTBoundary}). We start by describing the latter. Pick a curve ${\cal
C}$, parameterized as
$\left( \tau(u), z(u)\right) 
$, through $u \in [0,2\pi)$. We will be
interested in curves that are close to the original AdS boundary, so that $z\sim
\epsilon$. Let us note that the induced metric on the boundary curve is
\begin{equation*}
    ds^2_{\rm ind} = \frac{\tau'(u)^2 + z'(u)^2}{z(u)^2}du^2\,,
\end{equation*}
where primes denote derivatives with respect to the parameter $u$. One important
class of boundary conditions correspond to choosing the embedded curve to have
fixed proper lengh, $\beta$. For
this, pick a cutoff $\epsilon$ near the boundary, so that $\epsilon \rightarrow
0$. The fixed-length boundary condition then reads
\begin{equation*}
  \int_{0}^{2\pi}  \frac{\sqrt{\tau'(u)^2 + z'(u)^2}}{z(u)} du = \frac{\beta}{\epsilon}\, \qquad \Rightarrow \quad \tau(u) = f(u)\,,\quad z(u) = \epsilon \frac{2\pi}{\beta}f'(u) + {\cal O}(\epsilon^2)\,,
\end{equation*}
where we chose the arbitrary diffeomorphism $\tau(u) = f(u)$ and solved the
embedding constraint near the boundary, $\epsilon \ll 1$, and $z(u) = {\cal
O}(\epsilon)$.  The condition for this to be an orientation preserving map from
the circle boundary to itself requires that $f(u+2\pi) = f(u)+2\pi$ with $f'(u)
>0$. Geometrically, the cut-off Poincaré plane, endowed with the fixed-length
boundary is then described by the monotonic function $f(u)\in \mathrm{Diff}({\rm
S}^1)$. Physically, $f(u)$ will correspond to the remaining non-trivial degree
of freedom of the system, sometimes called a \textbf{boundary graviton mode}. As will
become clear in the following analysis, this mode has  the same
physical characteristics as the Goldstone mode in the SYK model, described in
Section \ref{sec:SYKspontaneous}, and also denoted $f$, for example in
\eqref{eq:ReparameterizationTransformation}. The boundary condition for the
dilaton at the cutoff surface reads  
$
   \Phi \sim \frac{\phi_r}{\epsilon} + {\cal O}(\epsilon),
$
where $\phi_r$ is a finite boundary field amplitude and the factor
$\epsilon^{-1}$ reflects the divergence of the bulk dilaton (cf.
Eq.~\eqref{eq:DilatonBulkSolution}), upon approaching the boundary. Let us note
that the boundary conditions we just described are left invariant by the action
of SL$\left(2,\mathbb{R} \right)$, implemented by
\begin{equation*}
  f(u) \longrightarrow  \frac{af(u) + b}{c f(u) + d}\,,\qquad \textrm{such that} \qquad ad-bc = 1\,,
\end{equation*}
which are inherited from the action of the bulk isometries on the boundary
curve. We conclude that the configuration space of nearly-AdS$_2$ solutions is
${\rm Diff} (S^1)/{\rm SL}(2,\mathbb{R})$\,.

\begin{BoxTypeA}[box:ads2-blackhole]{The AdS$_2$ black hole and its higher dimensional origins}

   A solution that illustrates several of the concepts we just introduced above
   is the 2D  black-hole. Using a convenient system of coordinates its
   Lorentzian line element reads
\begin{equation*}
ds^2 = -(r^2-r_h^2)dt^2 + \frac{dr^2}{r^2 - r_h^2}\,,\qquad \Phi(r) = \Phi_r r\,,
\end{equation*} 
This metric is locally related to Poincaré AdS$_2$. This can be seen via the coordinate transformation
\begin{align}
r = \frac{1}{z} \frac{1+\frac{r_h^2}{4} (t_P^2 + z^2)}{1-\frac{r_h^2}{4}(t_P^2 - z^2)}\,,\qquad t = \frac{2}{r_h} \arctan \left(  \frac{\frac{r_h}{2}t_P}{1+\frac{r_h^2}{4} (t_P^2 + z^2)} \right)\,,
\end{align}
which brings the metric back into the form of the Poincaré AdS$_2$ metric
Eq.~\eqref{eq:AdSMetricPoincare}. The parameter $r_h$ has the interpretation of
the horizon radius, as we will see in a few moments. While the metric is locally
related to Poincaré AdS$_2$, the dilaton has a non-trivial profile, whose value
at $r_h$ defines the entropy $S =  \frac{\Phi(r_h)}{4 G_N}$. Usually in
(semi-)classical gravity, black-hole entropy is related to horizon area of the
underlying black hole, instead here it is given by the value of the dilaton at
the horizon. (Furthermore, in two dimensions a black hole horizon is spatially a
point, so the notion of area doesn't really make sense a priori.) We can gain
more intuition on this expression of the entropy by exploring its
higher-dimensional origin, and at the same time also elucidate the relation of
JT gravity to its higher-dimensional cousins. Consider the four-dimensional
charged black-hole solution of mass $M$ and charge $Q$, given by the line
element
\begin{equation*}
ds^2 = - f(r) dt^2 + \frac{dr^2}{f(r)} + r^2 d\Omega_2^2\,,\qquad f(r) = 1-\frac{2M}{r} + \frac{Q^2}{r^2}\,.
\end{equation*}
This is a solution of four dimensional gravity, complemented by a Maxwell field,
$F_{rt} \sim Q/r^2 $, under which the hole is charged. The relevant
four-dimensional action reads 
 \begin{equation}\label{eq:EinsteinMaxwell}
        I_{4D} = \frac{1}{16\pi G_4} \int d^4x \sqrt{-g_4}\left( R_4 - \tfrac{1}{4}F_{\mu\nu}F^{\mu\nu}\right)\,.
\end{equation}
The condition for the function $f(r)$ to vanish is a quadratic, so there are two
horizons $r_\pm$. The entropy of this black hole is given by the celebrated
\keyword{Bekenstein Hawking formula}
\begin{equation*}
S = \frac{A}{4G_{(4)}} = \frac{\pi r_+^2}{G_{(4)}}\,,
\end{equation*}
where the area of the black-hole is determined by the so-called outer horizon,
$r_+$. If we take the charge --- taken positive for definiteness --- to
coincide with the mass $Q \rightarrow M$, we approach the so-called extremal
limit $r_+ \sim r_-$, which is characterised by a near-horizon geometry of the
form AdS$_2\times$S$^2$. The JT/Schwarzian theory emerges in this limit as the
universal description of the excitations above this extremal limit, and their
effective theory is precisely the theory of near-AdS$_2$ geometries furnished by
JT gravity as introduced above. Let us provide a few more details of how this
works.

Starting from the higher-dimensional Einstein-Maxwell theory
\eqref{eq:EinsteinMaxwell}, one makes the metric ansatz, $ds^2 = g_{\mu\nu}(x)
dx^\mu dx^\nu + \Phi(x)^2 d\Omega_2^2$, that is the four-dimensional metric
takes the form of a warped two-sphere over a general two-dimensional base
described by $g_{\mu\nu}$. This gives rise to a two-dimensional effective action
which already takes the form of a generalized dilaton theory, where $\Phi$ plays
the role of this dilaton. In order to land on the sought-after JT theory in two
dimensions, one further expands around the extremal limit $\Phi = \Phi_0 +
\phi$, with $\phi$ a small fluctuation around the extremal limit, and $\Phi_0 =
r_+$, then we obtain the effective action
\begin{equation*}
I_{4} \approx -S_0 \chi + \frac{\Phi_0}{2 G_{(4)}}\int d^2 x\sqrt{-g} \phi \left( R + \frac{2}{\Phi_0^2} \right) + \textrm{boundary}\,.
\end{equation*}
We first note that the standard near-extremal entropy of the 4D black hole
indeed shows up as the coupling to the Euler character, and that secondly we
simply need to identify $G_N = \frac{G_{(4)}}{8\pi \Phi_0} $ to recover the JT
action introduced at the beginning of this section. Finally, the Reader will no
doubt already have recognized that the 2D black hole solution corresponds
exactly to the dimensionally reduced near-extremal RN solution, where $r_h =
\Phi_0$, and that as advertised before the dilaton indeed plays the role of
transverse horizon area, once interpreted via this dimensional reduction.

In fact, JT and its Schwarzian formulation make an appearance as the effective theory of near-extremal AdS$_2$ solutions appearing in all sorts of higher-dimensional contexts, implying that they cover a universal subsector of the physics of higher-dimensional black hole horizons.
 
\end{BoxTypeA}

\subsubsection{Boundary diffeomorphisms and the Schwarzian action}

In this section, we discuss how the Schwarzian theory provides a powerful
connection between the SYK familiy of models and JT gravity. Having discussed
the emergence of the Schwarzian as an effective theory of the SYK model in
Section \ref{sec.SYK-Schwarzian}, we now turn to the JT side and start by noticing
that the bulk gravitational dynamics is trivial in the sense that the dilaton
equation of motion fixes the metric to be constant negative curvature, $R=-2$.
However, as we shall see now, we can view the shape of the boundary, described
by the reparametrization mode  discussed above in Section \ref{sec.ClassicalSols}
as a sort of boundary graviton mode\footnote{This terminology is borrowed from
the three-dimensional gravity literature, where the boundary gravitons furnish
Virasoro descendants [Henneaux type ref].} that gives non-trivial dynamics. We
recall from Section \ref{sec.ClassicalSols} that the `wiggly' boundary is fully
specified by the diffeomorphism $f(u)$, specifying the shape of the embedding of
the boundary into the cut-off Poincaré disc. By construction all such boundaries
have fixed proper length $\beta$, but different shapes have different action
costs, as we shall now establish. Evaluating the action on-shell, the only
non-trivial contribution\footnote{In discussing how the Schwarzian arises we can
neglect the topological term for the time being.} comes from the boundary term
\begin{equation*}
    I_{\rm bd} = -\int dx \sqrt{-\gamma} \Phi (K-1)\,,
\end{equation*}
for which we need to evaluate the extrinsic curvature term, $K$. Given outgoing unit normal $n^\mu$, this is defined as
\begin{equation}\label{eq:ExtrinsicCurvature}
K = \gamma^{ab} K_{ab}, \qquad K_{ab} = \gamma_a^\mu \gamma_b^\nu \nabla_\mu n_\nu
\end{equation}
where $\gamma_{ab}$ is the induced metric, and $\gamma^\mu_a = \partial x^\mu /
\partial \sigma^a$. In essence, this formula probes how the normal of the
curve covariantly changes, i.e. how `bent' the curve is relative to the straight
lines of the theory, its geodesics, cf. Fig.~\ref{fig:JTBoundary}. For the case in hand, the boundary is one-dimensional, so it
is parametrized by a single parameter $\sigma^a = u$, while in Poincaré
coordinates $x^\mu(u) = (\tau(u), z(u))$, as before. Calculating thus the
extrinsic curvature and expanding near the boundary -- that is to first order in
$\epsilon$, we obtain
 \begin{equation*}
        K = \frac{1}{\epsilon} \cdot \frac{1}{\tau'} + \epsilon\left( \frac{\tau'''}{\tau'^2} - \frac{3}{2}\frac{\tau''^2}{\tau'^3} \right) + O(\epsilon^3)\,.
    \end{equation*}
The first term is a $1/\epsilon$ divergence that is precisely cancelled by the
counterterm in the boundary action, while the ${\cal O}(\epsilon)$ term is
proportional to the Schwarzian derivative, \eqref{eq:SchwarzianAction}. Putting
it all together, we find that the JT action has reduced to the Schwarzian
 \begin{equation*}
        I_{\text{Sch}} = -\phi_r \int_0^\beta du\, \{\tau(u), u\}\,,
    \end{equation*}
i.e. the gravitational analaog of Eq.~\eqref{eq:SchwarzianAction}.

Let us summarize what we have learned about the classical behavior of JT gravity
thus far. Firstly, the  theory is a stand-alone two-dimensional toy model
of Einstein gravity that gets around the triviality of vanilla Einstein-Hilbert
theory in this dimension by adding a further degree of freedom, the dilaton.
Furthermore, the exact form of the JT action can be derived via an s-wave
reduction of four-dimensional near-extremal black holes, whence we understand JT
to provide a powerful effective theory describing the dynamics of the
near-infinite AdS$_2$ throat and its excitations. As such JT gravity plays an
important role in the dynamics of more physical, higher-dimensional black holes,
a connection that was fruitfully explored in recent literature
\cite{Iliesiu:2020qvm,Heydeman:2020hhw}. Classical solutions are described by a wiggly disc cut out of
the constant negative curvature hyperbolic plane. This finite cutoff boundary
breaks the full asymptotic symmetry group of the hyperbolic disc
\begin{equation}
    \textrm{Diff}(S^1) \longrightarrow \textrm{SL}(2,\mathbb{R})
\end{equation}
and we can understand the Schwarzian mode as the Nambu-Goldstone boson of this
    symmetry breaking. Consequently, the Schwarzian, giving the leading action
    cost for this breaking, is the corresponding soft-mode action, in a pattern
    that perfectly mirrors the SYK story as shown in \ref{sec:SYKspontaneous}.
    As a final comment, let us note that the Schwarzian itself gives a
    non-trivial action cost for a non-trivial diffeomorphism $f\in $
    Diff($S^1$), but is left invariant for $f \in $ SL$(2,\mathbb{R})$.

    \subsubsection*{The Schwarzian path integral on the disk and
    trumpet}\label{sec:DiskTrumpetIntegral} 
    
\begin{figure}[h]
\centering
\includegraphics[width=0.4\textwidth]{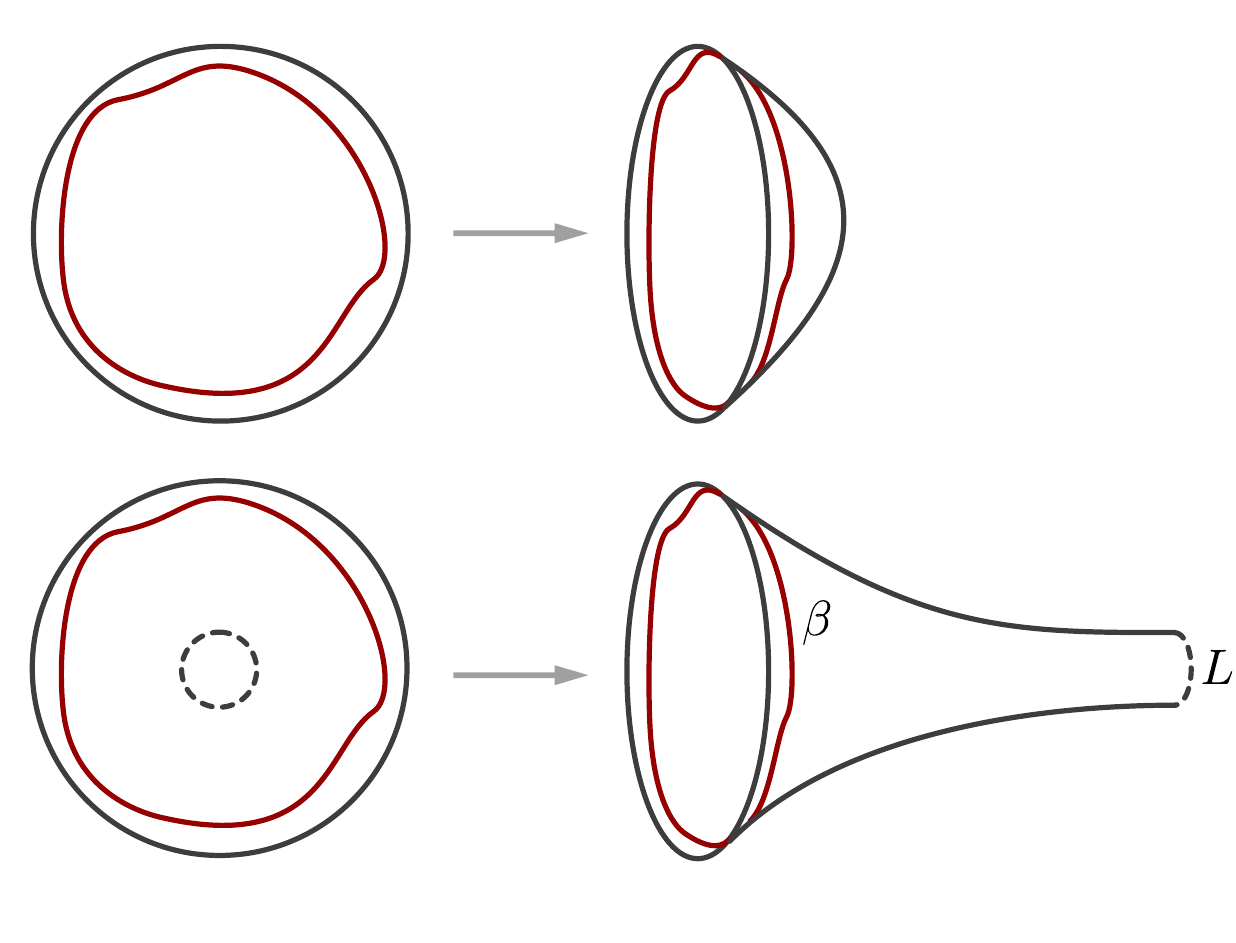}
\caption{The top row shows the disk topology with the wiggly asymptotic boundary in red. The Schwarzian path integral on this geometry is one-loop exact and its explicit evaluation is described in the main text. The bottom row shows the so-called `trumpet' geometry, which has two boundaries and thus has the topology of an annulus. The red boundary is the same kind of asymptotic wiggly boundary as above, while the inner, dashed, boundary is geodesic. The path integral again is one-loop exact and is performed in the main text. In both cases, the representation on the right embeds these geometries in three dimensions, which is useful when visualising certain gluing constructions that will be employed to analyse the JT path integral further below.}
\label{fig:Trumpet}
\end{figure}

    In preparation for the next section, let us discuss the quantum behavior of
    the Schwarzian on two special geometries, which both play central roles in
    the development of the theory.  The first geometry in question is simply
    the hyperbolic disc, from which we deduce the leading spectral density of
    the Schwarzian theory. The second geometry is the so-called trumpet
    geometry, which topologically is an annulus whose outer boundary is
    asymptotically hyperbolic, while the inner boundary is geodesic, cf.
    Fig.~\ref{fig:Trumpet}. 

    Executing the boundary mode path integral for these two geometries is an important
    step towards the expansion in general bulk topologies and the  quantization
    of the JT theory in Section \ref{sec:TopoJT}. At the same time, the details
    of the computation are somewhat technical (cf. Box~\ref{box:disk-trumpet}), and we here restrict ourselves to stating the result, the JT partition
    sums of disk and trumpet after integration over boundary graviton modes 
    ~\cite{Saad2019,StanfordWitten2017Schwarzian}\phantomsection\label{sec:trumpetIntegral}
\begin{equation}\label{eq:Zdisk}
    Z_{\text{disk}}(\beta) = e^{S_0}\,\frac{\phi_r^{3/2}}{(2\pi)^{1/2}\,\beta^{3/2}}\; e^{\frac{2\pi^2 \phi_r}{\beta}}\,,
\end{equation}
\begin{equation}\label{eq:Ztrumpet}
    Z_{\mathrm{tr}}(\beta, L) = \frac{\phi_r^{1/2}}{(2\pi\beta)^{1/2}}\; e^{-\frac{\phi_r L^2}{2\beta}}\,.
\end{equation}
In both cases the exponential factors are simply the classical saddle-point actions, while the prefactors come from a one-loop determinant calculation.
Conveniently, the disk result \eqref{eq:Zdisk} gives the leading density of states via an inverse Laplace transform, 
\begin{equation}
    \label{eq:SpectralDensityJT}
  \boxed{  \rho_0(E) \sim e^{S_0} \sinh(2\pi\sqrt{2\phi_r E})}
\end{equation}
 which is the celebrated Schwarzian spectral density. The trumpet partition
 function \eqref{eq:Ztrumpet} plays the role of a `propagator' connecting
 asymptotic boundaries to internal geodesic boundaries, and will be a key
 building block in the topological expansion of the full JT path integral in Section
 \ref{sec:TopoJT}. Before going there, it will be helpful to first review random
 matrix theory, paying special attention to a set of quantities and
 constructions that will naturally appear in JT.

\begin{BoxTypeA}[box:disk-trumpet]{Boundary gravitons and the Schwarzian path integral on the disk and
    trumpet}

To set up the path integral computing the thermal trace of the Schwarzian, we
first change variables from $\tau(u)$ to $t(u)$ via $\tau = \tan(t/2)$, with
$t(u+\beta) = t(u) + 2\pi$, which is equivalently a field redefiniation mapping
the circle to a line. The path integral in the $t(u)$ variable is a convenient
choice for the computation of the disc partition function, because the measure
is simply the naive flat choice. 
Using the composition rule for the Schwarzian derivative,
the disk action becomes
\begin{equation}\label{eq:DiskActionCircle}
    I_{\text{disk}}[t] = -\frac{2\pi\phi_r}{\beta} \int_0^{2\pi} du\, \left( \{t,u\} + \tfrac{1}{2} t'^2 \right)\,,
\end{equation}
where the extra $\frac{1}{2}t'^2$ piece arises from the chain rule. In these
variables $t \in \mathrm{Diff}(S^1)$, and the SL$(2,\mathbb{R})$ isometry of the
Poincar\'e disc acts as $t \to \frac{a\tan(t/2)+b}{c\tan(t/2)+d}$. Since, as
above, this is a gauge redundancy, the path integral runs over
$\mathrm{Diff}(S^1)/\mathrm{SL}(2,\mathbb{R})$. Let us remark that SL$(2,\mathbb{R})$ appears as a global symmetry in SYK, and the fact that it is gauged in the bulk is an example of a general feature in holographic duality, namely that global symmetries of the boundary appear as gauge symmetries in the bulk. 

For the trumpet, one instead takes the quotient of the upper half-plane by the
hyperbolic transformation $z \sim e^b z$, producing a cylinder with one
asymptotic and one geodesic boundary of length $L$. The isometry group is
reduced to $\mathrm{U}(1)$, which geometrically corresponds to rotations of the
annulus (or cylinder) defining the trumpet. A similar boundary analysis to the
case above, yields the trumpet action $I_{\text{tr}}$ identical to the disk
action, except for the replacement of the $\frac{1}{2}t'^2$ term by
$-L^2 t'^2/2(2\pi)^2$, 
and the path integral now running over $\mathrm{Diff}(S^1)/\mathrm{U}(1)$, with the U$(1)$ rotation gauged. A
remarkable property of both path integrals is that they are one-loop exact: the
saddle-point approximation discussed above gives the full all-loop answer \cite{StanfordWitten2017Schwarzian}, meaning to arbitrary perturbative order in the Schwarzian coupling $\phi_r$.  This can be understood as a consequence of
the Duistermaat--Heckman theorem, since the field spaces
$\mathrm{Diff}(S^1)/\mathrm{SL}(2,\mathbb{R})$ and
$\mathrm{Diff}(S^1)/\mathrm{U}(1)$ carry natural symplectic structures, and the
Schwarzian action itself is the moment map for the $S^1$ action $t(u)\mapsto
t(u+\theta)$. The partition function then localizes to the unique fixed point of
this action, namely the classical solution $t_{\text{cl}}(u) = u$, giving
\begin{equation}\label{eq:DHFormula}
    Z = \frac{e^{-I[t_{\text{cl}}]}}{\sqrt{\det H}}\,,
\end{equation}
where $H$ is the generator of the $S^1$ action, $H =
\frac{i\phi_r}{\beta}\partial_u$, with 
eigenvalues $\frac{\phi_r}{\beta}
n$, $n\in \mathbb{Z}$ in the Fourier basis $e^{inu}$. For the disk, the modes $n = -1, 0, 1$ lie in the
$\mathfrak{sl}(2,\mathbb{R})$ stabilizer and are excluded, giving
\begin{equation}
    \det(H)_{\text{disk}} = \prod_{|n|\geq 2} \frac{\phi_r}{\beta}\, n = 2\pi\left(\frac{\beta}{\phi_r}\right)^3\,.
\end{equation}  
For the trumpet, just $n=0$ is excluded as it lies in the
$\mathrm{U}(1)$ stabilizer. Using $\zeta$-function regularization: $\prod_{n\geq 1} n = \sqrt{2\pi}$ and
$\prod_{n\geq 1} ({\phi_r}/{\beta}) = ({\phi_r}/{\beta})^{\zeta(0)} =
(\beta/\phi_r)^{1/2}$, the product evaluates to $\det(H)_{\text{tr}}
=\frac{2\pi\beta}{\phi_r}$.
We finally consider the classical action at $t_{\text{cl}}(u) = u$, to find
$-I[t_{\text{cl}}]_{\text{disk}} = \frac{2\pi^2 \phi_r}{\beta}$ and
$-I[t_{\text{cl}}]_{\text{tr}} = -\frac{\phi_r L^2}{2\beta}$.
Putting the pieces together, and including the topological term $e^{S_0}$ (with
$S_0 = 2\pi\Phi_0$, and $\chi = 1$ for the disk, $\chi = 0$ for the
trumpet), we obtain Eqs.~\eqref{eq:Zdisk} and \eqref{eq:Ztrumpet}. 
\end{BoxTypeA}

\section{Matrix Theory}
\label{sec:MatrixTheory}
Alongside the actual bulk and boundary systems, \textit{matrix ensembles} play a
central role as physical proxies in the holographic framework. Historically,
ensembles of  matrices have been an important element of theory building, both in
quantum chaos and in high-energy physics/gravity. However, the two fields have
been looking at these structures from different perspectives: employing
matrix ensembles to the phenomenological description of 
bulk geometries or the modelling of gauge field correlations, the latter has been primarily concerned with the topological
classification of correlation functions. By
contrast, quantum chaos puts emphasis on the spectral and
eigenfunction statistics of matrix Hamiltonians. While these two perspectives
have been developing largely independently of each other, the recent advent of
holographic dualities has brought them together, revealing structures that were
previously not known, or at least underappreciated.

To make all this more concrete, let us consider the two-point correlation
function
\begin{align}
    \label{eq:MatrixCorrelationFunction}
    R_2(\omega)\equiv \left\langle \rho(\omega/2) \rho(-\omega/2) \right\rangle_{\mathrm{c} },
\end{align}
where $\rho(E)$ is the spectral density of a Gaussian distributed random
matrix Hamiltonian $H$ with variance $\left\langle H_{\mu\nu}H_{\rho \sigma}
\right\rangle=\frac{\lambda^2}{N}\delta_{\mu \sigma}\delta_{\nu \rho}$ (the
Gaussian unitary ensemble, or matrix ensemble of class A), and
$\langle \dots \rangle_{\mathrm{c} }$ is the connected average. From random
matrix theory, we know that this correlation function equals
\begin{align}
    \label{eq:RMTCorrelationFunction}
    R_2(\omega)=-\rho^2 \frac{\sin^2(s)}{s^2},\qquad s\equiv \pi \rho \omega, \qquad s\not=0, 
\end{align}
where $\rho=N/\pi \lambda$ is the average spectral density at the center of the
spectrum. Its frequency Fourier transform, is the  \keyword{spectral form
factor},\footnote{The shift factor $-1$ in this definition is due to our
exclusion of the singular auto-correlation of levels for $\omega=0$.} 
\begin{align}
    \label{eq:SpectralFormFactor}
    K(t)=\frac{1}{\pi \rho}\left(\frac{t}{t_\textrm{H}}-1  \right)\Theta(t_\textrm{H}-t ), 
\end{align}   
 with its notorious `ramp' followed by a `plateau' for times exceeding the
 \keyword{Heisenberg time}, $t_\mathrm{H}=2\pi \rho$.

 \begin{figure}[t]
\centering
\includegraphics[width=0.8\textwidth]{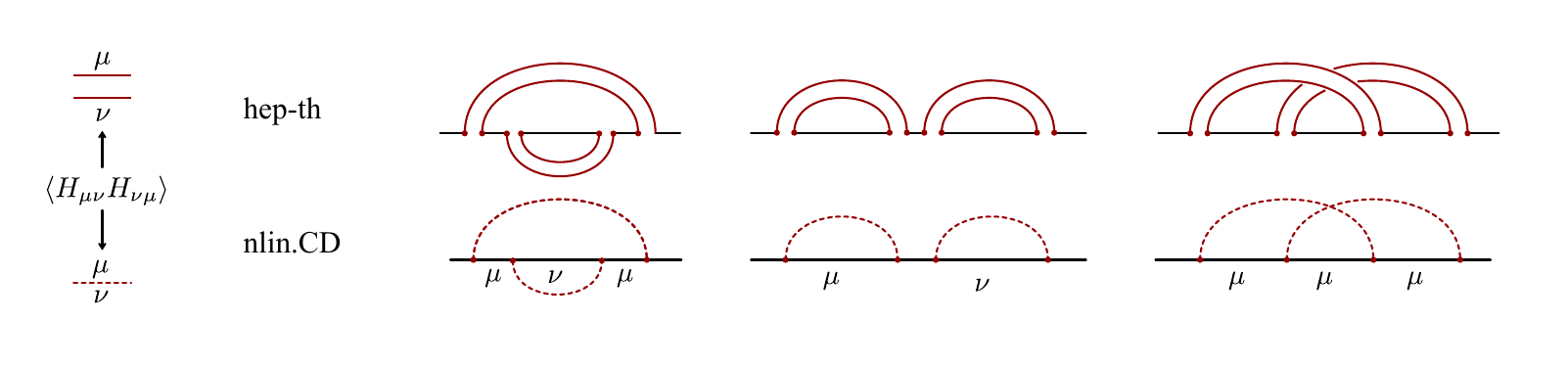}
\caption{Top: Ribbon diagram representation of matrix contractions (left) and
fourth order expansion of the matrix Green function (right). Bottom: Impurity
diagram representation of the same objects. }
\label{fig:RibbonVsImpurityDiagrams}
\end{figure}

\begin{BoxTypeA}[box:ribbon-diagrams]{Ribbon diagrams vs. impurity diagrams}

    Particle physics and condensed matter physics have independently developed
diagrammatic codes representing the perturbative expansion of matrix correlation
functions. However, reflecting their different foci, these  languages
are  somewhat different. In particle physics, it is customary to represent the
Gaussian contraction of pairs of matrix elements in terms of ribbons, as shown
in the upper left part of Figure \ref{fig:RibbonVsImpurityDiagrams}. The upper
right part exemplifies this notation for the matrix Green function $G(z)$
expanded to fourth order in $H$. In this notation `non-planar' diagrams with
finite genus degree contain crossing ribbons. 

Condensed matter physics and quantum chaos represent matrix contractions  by
single dashed lines, see the bottom part of the figure. Here, non-planarity
translates to the presence of crossing lines. This notation is more efficient in
infinite order resummations, and we will use it in the following. Note that
intersecting dashed lines, equivalent to non-planar ribbon diagrams, come with a
price: for each intersection, we loose one free running index summation, and
hence a factor $N$.
\end{BoxTypeA}

\begin{figure}[h]
\centering
\includegraphics[width=0.8\textwidth]{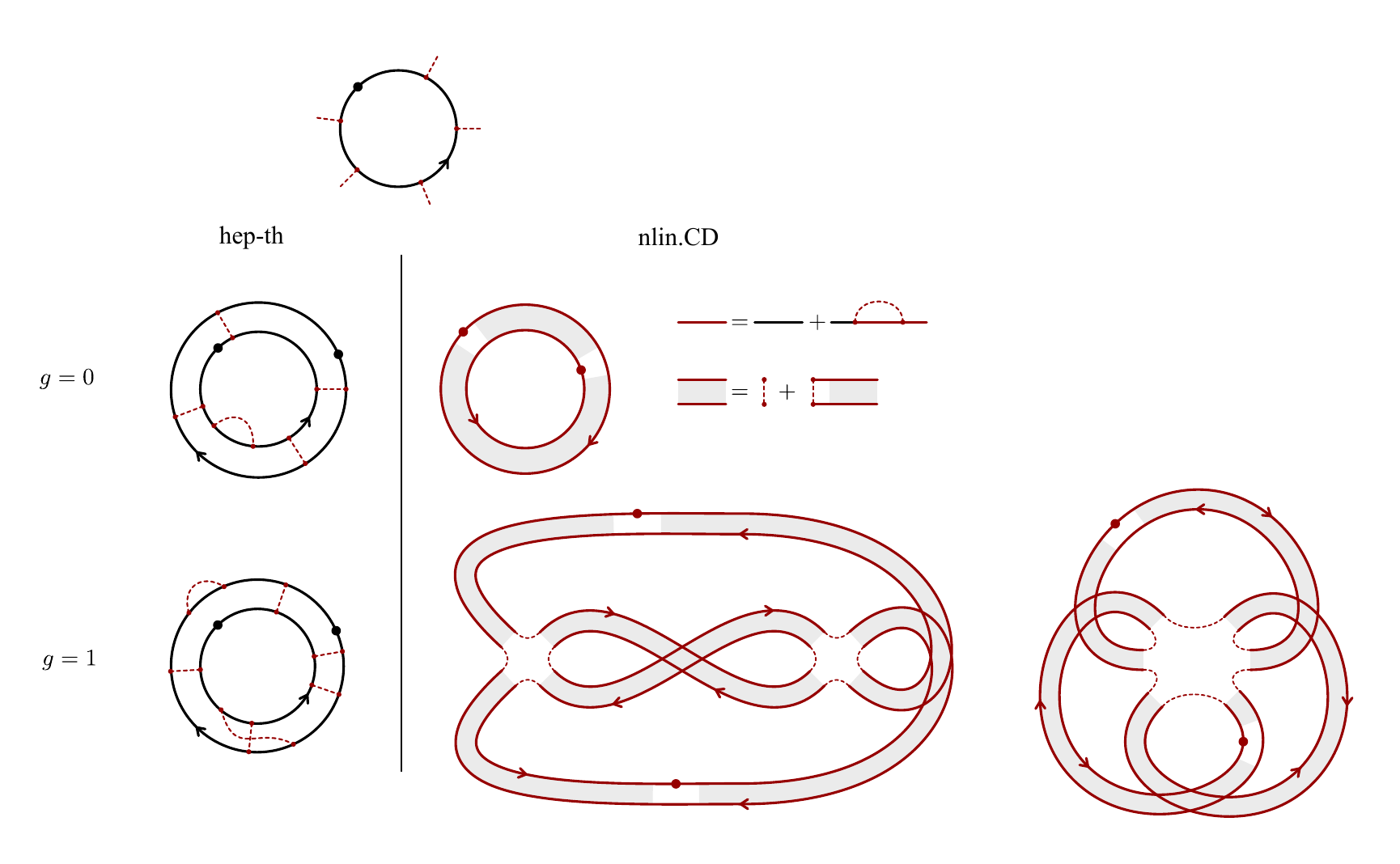}
\caption{Graphical representation of the trace of the random matrix resolvent
(closed solid line) expanded to fourth order in $H$ (dashed lines).
Left: the topologically oriented particle physics perspective of averaged
correlation functions classifies them according to their genus, as quantified by
the number of intersecting dashed lines. Right: the quantum chaos perspective
up-emphasizes the singularity of correlation functions in small parameter
differences, which generally requires infinite order resummations. Further
discussion, see text. }
\label{fig:GenusExpansion} 
\end{figure}

Suppose we didn't know the answer Eq.~\eqref{eq:RMTCorrelationFunction}
beforehand, and wanted to compute it from scratch. A naive approach might
start from a representation of the spectral density in terms of Green functions,
$\rho(E)=-\frac{1}{\pi} \mathrm{Im}\,\mathrm{tr} (G(E^+)) $, where $G(z)=(z
-H)^{-1}$, and expand in $H$ to obtain a formal series representation
diagrammatically represented as in Fig.~\ref{fig:GenusExpansion}, top, where the
black line segments denote the zeroth order propagator, $1/z$,  dashed lines
matrix elements $H_{\mu\nu}$, and the closure of the ring reflects the trace operation.

\subsection{Particle physics perspective: topological recursion}

Subsituting this expansion into Eq.~\eqref{eq:MatrixCorrelationFunction}, we
obtain a double-ring structure where averaging over $H$ enforces pairwise
contractions of dashed lines. Particle physics generally looks at the resulting
diagrams --- see Fig.~\ref{fig:GenusExpansion}, left for two examples --- through the
lens of topology: each diagram can be associated with a genus degree, $g$.
In  the double line representation  of matrix diagrams (see Box~\ref{box:ribbon-diagrams}), this degree
equals the Euler characteristic of the surface tilings defined by patterns of
ribbons. Equivalently, we can ask: what is the minimal genus of a
surface on which a given diagram can be drawn without intersecting dashed lines?
For example, the diagram on the top left is obviously planar, $g=0$, while the one on
the bottom left has $g=1$: it can be drawn without intersections on a torus. 

This topological classification is more than a bookkeeping
instrument. We had already noted that each intersection of dashed lines
translates to the loss of a free running index summation, and hence a factor
$N$. The genus classification therefore organizes the series in powers of $1/N$.
With $R(z)=\tr(G(z))$, we can formulate this statement as
\begin{align}
    \label{eq:TopologicalExpansion}
    \left\langle R(z_1) R(z_2)\dots R(z_n) \right\rangle_{\mathrm{c}H }\equiv \sum_{g} \frac{R_{g,n}(z_1,\dots,z_n)}{N^{2g+n-2}}.
\end{align} 
Of course, this expansion remains formal unless we can say something about its
coefficients $R_{g,n}$. We extract this piece of information from an exact
representation of the random matrix average in an eigenvalue representation: For
a random matrix ensemble with probability density $P(H)\propto \exp(-N\,
\mathrm{tr} V(H))$,  where $V(H)$ is a unitarily invariant weight function
$V(H)=V(UH U^\dagger)$ ($V(H)=\frac{1}{\lambda^2} \,\tr(H^2) $ for the GUE), and
equally invariant observables $\mathcal{O}(H)$, we have~\cite{mehtaRandomMatrices1991}
\begin{align}
    \label{eq:EigenvalueRepresentation}
    \left\langle \mathcal{O}(H) \right\rangle =
    \frac{1}{Z} \int  D\lambda\, e^{-N  V(\lambda)} \Delta^2(\lambda) \mathcal{O}(\lambda)\equiv \left\langle \mathcal{O}(\lambda) \right\rangle ,
\end{align}
where $\lambda=(\lambda_1,\dots,\lambda_N)$ are $H$'s eigenvalues, $D
\lambda=\prod_i d \lambda_i$,  $\Delta(\lambda)=\prod_{i<j} (\lambda_i -\lambda_j)$ is
the
Vandermonde determinant, and $Z$ a normalization constant.

Eq.~\eqref{eq:EigenvalueRepresentation} introduces two concepts which will be
individually key to our later discussion: topological recursion, and the spectral
curve. Both are introduced via the engineering of `intelligent zeros',
based on Eq.~\eqref{eq:EigenvalueRepresentation}: For any function
$\mathcal{X}(\lambda)$, we have
\begin{align*}
    0 = \frac{1}{Z} \int  D\lambda\, \partial_{\lambda_i}
     \left(\mathcal{X}(\lambda) \Delta^2(\lambda) e^{-N  V(\lambda)} 
     \right),
\end{align*} 
i.e. the unbounded integral of a total derivative vanishing.
The trick now is to integrate by parts and in this way represent the zero as a
sum of individually non-trivial terms. Equations generated in this way are
called \keyword{loop equations}. The art of loop equations is to come up with
choices of $\mathcal{X}(\lambda)$ such that the resulting equations contain
contributions to the correlation function hierarchy of different order $n$
and/or genus degree. 

As often in random matrix theory, the design of these systems of equations is
motivated by formal reasoning and may not be particularly inspiring for the
uninitiated reader. However, to illustrate the principle, consider the simplest case
$\mathcal{X}(\lambda)=\frac{1}{z-\lambda_i} $. It is
straightforward to demonstrate that the insertion $\sum_i
\partial_{\lambda_i}\left( X(\lambda) \dots\right)$ yields the loop equation 
\begin{align}
    \label{eq:LoopEquationOnePoint}
    \left\langle R(z)^2  \right\rangle = -N \left\langle \tr\left(     \frac{V'(H)}{z-H}  \right)  \right\rangle. 
\end{align} 
We can now expand this exact equation in $1/N$. For example, to leading order,
$\left\langle R(z)^2  \right\rangle \approx \left\langle R(z)
\right\rangle^2\approx (N R_{0,1}(z))^2 $. The appearance of single averaged
resolvents in the large $N$ limit, $R_{0,1}(z)$, in this expression opens an interesting
avenue, namely the analytic characterization of spectral densities by the concept
of the so-called \textbf{spectral curve}. To define the latter, we play
another mathematical trick and add $V'(z)^2/4$ to both sides of the equation. It
is then straightforward to massage it into the form 
\begin{align*}
    \left( R_{0,1}(z)-\frac{V'(z)}{2}  \right)^2 = 
    \frac{V'(z)^2}{4}-  \frac{1}{N} \left\langle \tr \left( \frac{V'(z)-V'(H)}{z-H}  \right) 
     \right\rangle\equiv P(z),
\end{align*} 
where the average on the right-hand side is to be evaluated to leading order in
$1/N$. Assuming a polynomial weight, $V(z)$, the right-hand side is a
polynomial, $P(z)$, and this its only feature we will need in the following.
Defining 
\begin{align}
    \label{eq:SpectralCurveDefinition}
    y(z)\equiv R_{0,1}(z)-\frac{V'(z)}{2},
\end{align}
we can rewrite the above equation as $y(z)^2=P(z)$.
This quadratic equation  has two solutions, $\pm y(z)$; the branch $y(z)$
defined by Eq.~\eqref{eq:SpectralCurveDefinition} is what we call the spectral
curve. The presence of the resolvent $R_{0,1}(z)$ on its right-hand side
heralds a connection between
the spectral curve and the system's spectral density, and this is what makes it
interesting to us. To substantiate this point, note that for two closeby points
$z=E\pm i0$, we have the options $y(E+i0)\approx \pm y(E-i0)$. Now observe that $y(z)$
equals the averaged resolvent, plus a harmless polynomial. Inside regions of
support of the spectral density, $\rho(E)\approx - \frac{1}{2\pi i}
(R_{0,1}(E+i0)-R_{0,1}(E-i0))$, the function $y(z)$ must have a discontinuity,
i.e. $y(E+i0)=-y(E-i0)$. At the same time,
$y(E+i0)-y(E-i0)=R_{0,1}(E+i0)-R_{0,1}(E-i0)$. Combining these equations we
obtain  
\begin{align}
    \label{eq:SpectralDensityFromSpectralCurve}
y(E\pm i0 )= \mp i \pi \rho(E) . 
\end{align}

\begin{BoxTypeA}[box:spectral-curve]{Why do we care about the spectral curve?}

Eq.~\eqref{eq:SpectralDensityFromSpectralCurve} shows how the spectral curve is
determined by the spectral density, the latter being an observable  of physical
significance. Why, then, are we interested in the seemingly more abstract
spectral curve? From the perspective of abstract matrix theory, the merit of the
spectral curve is that it 'analytically continues' information carried by the
spectral density on the real axis into the complex plane. This is important, in
particular in the vicinity of spectral edges, where much of the system's physics
is conditioned by analytic properties of the spectrum. Another motivation
follows from the holographic context, where we note that all our present
discussion is perturbative in $N$ (we are talking about a genus/large $N$
\textit{expansion}). However, eventually, we will be ambitious to push the
description of the gravitational bulk towards single level resolution. As we
will discuss in Section~\ref{sec:UniverseFieldTheory}, this extension requires
elements of topological string theory whose construction builds on the spectral
curve as a backbone. 
\end{BoxTypeA}

In simple cases --- one-cut  matrix models with polynomial $V(H)$ --- the spectral curve can be defined directly
from the planar density by introducing an analytic function $y(z)$ whose
boundary values satisfy Eq.\eqref{eq:SpectralDensityFromSpectralCurve}. For
example for the Wigner semicircle, $\rho(E)=\frac{1}{2\pi \lambda^2}
\sqrt{E^2-4 \lambda^2}$, we obtain $y(z)=\frac{i}{2 \lambda^2} \sqrt{z^2- 4
\lambda^2}$ just by substitution $E\to z$. For the discussion of
more complex cases, we refer to the literature, e.g. Ref.~\cite{eynardRandomMatrices2018}.

\textit{Topological recursion:} Turning back to the loop equation Eq.~\eqref{eq:LoopEquationOnePoint}, we may
push the expansion to next-to-leading order in $1/N$. While the algebra becomes
more complicated, the emergence of a recursive structure is
relatively straightforward to see. For example, at next leading order we
encounter both, terms $\propto R_{1,1}(z)R_{0,1}(z)$ (i.e. one of the two
factors $R(z)$ on the left-hand side of Eq.~\eqref{eq:LoopEquationOnePoint} expanded to
next-to-leading order), and terms $\propto R_{0,2}(z,z)$ (i.e. the leading statistically
connected contribution to $\langle R(z)^2 \rangle$). In this way, a recursive
hierarchy involving a non-planar contribution to the one-point function and one of
lower order to the two-point function emerges. At the same time, the actual
execution of the program, even for the one-point function, may look hopelessly
complicated. It is therefore all the more remarkable that Eynard and
Orantin~\cite{Eynard:2007kz} managed to formulate recursion relations for the expansion
coefficients of arbitrary order $n$ and genus $g$. For our purposes, the precise
form of these relations, which look expectedly complicated, is not important.
What maters is that a `code book' describing the expansion of invariant matrix
ensembles subject to specific probability weights (equivalently spectral curves)
has been derived. As we will see in the next Section, its  highly structured data provides a testbed for the equivalence
of differently defined models, such as gravity, to matrix ensembles.

\begin{BoxTypeA}[box:double-scaling]{Double scaled and doubly non-perturbative matrix theory}

The field of holography appears to be  fond of the attribute `double' and uses
the term in at least three different contexts relevant to our present
discussion. The first is the \textbf{double-scaled SYK model}, previously
introduced in Section~\ref{sec:ChordDiagrams}. Second, we already mentioned that
the holographic correspondence probes near ground state energies of all three,
boundary and bulk theory, and their matrix model proxies. At the same time, we
often want to benefit of semiclassical principles, i.e. stay clear of the ultra
quantum regime with single level resolution. In practical terms, this means
testing energies $E$ asymptotically small compared to the bandwidth, $E/\lambda
\to 0$, but at the same time scaling $N\to \infty$ in such a way that there
remain many levels below  us. With a near edge spectral density $\rho\sim N
E^{1/2}\lambda^{-3/2}$, the number of levels in the $E$-interval scales as $N_E
\sim N (E/\lambda)^{3/2}$, so that the double-scaling limit requires keeping the
combination $E N^{2/3}$ large when sending $E \to 0$. This way of zooming in
defines the \keyword{double scaled matrix model}. (These relations hold for
 models whose near edge spectral curve reduces to the Airy curve, $y^2\sim
z$, as is the case for matrix polynomials $V(H)$ containing leading Gaussian
contributions. For models with more exotic weight functions, different profiles
of the near edge spectral density and level spacing emerge.) 

In the gravitational context, and in SYK physics, $N\sim \exp(S)$, is exponential
in the system's entropy. A $1/N$ expansion thus is non-algebraic in $S^{-1}$,
and  called (singly) non-perturbative. In this reading, probing physics at
level spacing scales, e.g. resolving the $\sin^2(\omega\times N/\pi \lambda)$
oscillations in the two-point function non-algebraic in $N\sim \exp(S)$, means
entering a \keyword{doubly non-perturbative} regime.

The discussion above assumes a generic (non-critical) spectral edge, where
universality guarantees the square-root density $\rho \sim \sqrt{E}$ regardless
of the matrix potential. This is the situation relevant to JT gravity and the
SYK model. More generally, one may fine-tune the matrix potential to a
multi-critical point, where the edge density changes to $\rho \sim E^{m-1/2}$
with $m = 2, 3, \ldots$, leading to quantitatively different double-scaling
limits with $E \sim
N^{-2/(2m+1)}$~\cite{Douglas:1989ve,Brezin:1990rb,Gross:1989vs}. These
multi-critical models describe the $(2,2m{-}1)$ minimal string theories and have
played a central role in the non-perturbative study of two-dimensional quantum
gravity; see Box~\ref{box:jt-matrix-model} for further discussion.
\end{BoxTypeA}

\subsection{Quantum chaos perspective: singular correlations}
\label{sec:singularCorrections}
In  quantum chaos, the topological contents of the correlation function
Eq.~\eqref{eq:MatrixCorrelationFunction} essentially is a non-issue (although we
will see how it creeps in through the backdoor). Instead, the focus is  on the
study of singularities arising in the limit of small energy differences
$\omega$. The exact random matrix result, Eq.~\eqref{eq:RMTCorrelationFunction},
exemplifies these singularities via the $\sim (\rho\omega)^{-2}$ power law
asymptotics. More generally, the singular behavior of correlation functions such
as $\left\langle R(E+ \omega/2 +i0) R(E- \omega/2 -i0) \right\rangle $   follows
from the presence of causal symmetry breaking, cf. the discussion below
Eq.\eqref{eq:GaussianFunctional}. In it,  $\omega$ plays the role of an explicitly
symmetry breaking parameter (akin to a magnetic field in a ferromagnet).
Symmetry breaking being a large $N$-phenomenon, this parameter must enter in the
dimensionless combination $\omega \times N /\lambda\sim \pi \omega \rho\equiv
s$, and correlation functions are expected to show power law singularities in
it. 

The
proportionality  $s\propto N$ indicates that the singularity degree $\sim
s^{-l}$ of a correlation function must be linked to its genus degree $g$.
However, note that the topological
expansion in $1/N$ per se is oblivious to a second expansion parameter $1/s\sim
1/(\epsilon N)$. Crucially, $s$ can become small even when $N$ is large,
indicating the presence of  regimes where high-order
topological expansion, or even non-perturbative analyses are required to
capture the relevant contributions. This is
precisely what happens, as we venture into the `plateau regime', probing energy scales of the order of the level
spacing $\omega \sim \Delta$, or $s\sim 1$. 

In perturbation theory,  competitions between a small symmetry breaking
parameter and a naive expansion in $1/N$ generally show at the level of infinite
order resummations. Fig.~\ref{fig:GenusExpansion}, top right exemplifies this
principle for the planar diagram class. The infinite order (in $H$) diagrams
shown there sum over all contributions without intersecting dashed lines. These
include the red single line propagators dressed by  non-crossing `self energy'
diagrams (defined by the indicated Dyson equation), and shaded gray area
two-propagator ladder diagrams (defined by the indicated Bethe-Salpeter
equation). Each of the latter, contributes a factor
$\sim 1/s$, leading to the overall planar contribution $\sim s^{-2}$, i.e. the
leading, and only power law dependence of the two-point function. 

At higher orders in the $s^{-1}$-expansion the analysis
gets expectedly more complicated. With a little imagination one
can see that  the 
diagram shown in the bottom right of Fig.~\ref{fig:GenusExpansion} can be drawn
without dashed line intersections on a torus: $g=1$. However, to actually
compute  the
matching $s$-power, one needs to represent the intersection crosses as sums of
ladders (this time connecting Green functions of identical causality $++$ or
$--$) and demonstrate that they add up to factors $\sim s$. With $6$
light-shaded ladders and two crosses, we obtain the expected power law $s^{-4}\sim N^{-(2g+2)}$
expected for a genus $g=1$ diagram. The reason why this power does not feature in the
overall result Eq.~\eqref{eq:RMTCorrelationFunction} is that the diagram gets
cancelled by the one shown on the right. (As a consistency check, note that with
five ladders and one vertex we get the same $s^{-4}\sim s^{-5+1}$.)

Summarizing, the broad class of genus $g$ diagrams contains a subclass of highly
regular ladder structures whose summation yields contributions that not only
scale as $N^2 N^{-(2g+2)}$ but also as $\omega^{-(2g+2)}$, leading the overall
scaling $N^2 s^{-(2g+2)}$. The identification of these contributions in the
JT partition sum will be proof of the statement that the latter describes
chaotic correlations, at least in perturbation theory. However, before turning
to this discussion it is worth noting that the identfication of chaos (in
gravity) will make essential reference to both, the topological degree of
individual contributions to correlation functions, and their  $s$-singular
subclasses. The above discussion highlights that the manual resummation of these
contributions is impractical. However, as we review in Box~\ref{box:sigma-model}
below, the nonlinear $\sigma$-model provides
us with an efficient tool for bridging between topology and singularity in
perturbation theory, and for the subsequent extension into the non-perturbative
regime: 

\begin{BoxTypeA}[box:sigma-model]{Nonlinear $\sigma$-model}

    Previously, we had introduced the nonlinear $\sigma$-model
    Eq.~\eqref{eq:ActionCollectiveFluctuations} as the effective action of the
    SYK model at large time scales. A simpler variant of this action 
    \begin{align}
        \label{eq:ActionSigmaEpsilon}
        S[Q]=-   i\pi \rho \,\mathrm{str} (Q \hat \epsilon),
    \end{align} 
    describes ensemble averaged correlations of the class A Gaussian random
    matrix ensemble (see Ref.~\cite{altlandQuantumChaosEdge2024} for the
    derivation)  around a characteristic energy $E$. Here, $\rho= \left\langle \rho(E)
    \right\rangle $ is the average spectral density at $E$, and $\hat
    \epsilon=\textrm{diag}(\epsilon^{+\textrm{b}},
    \epsilon^{-\textrm{b}},\epsilon^{+\textrm{f}},\epsilon^{-\textrm{f}})$ a
    diagonal matrix containing probe energies $\epsilon^{\pm \textrm{b/f}}\simeq
    E$. As with the theory defined by
    Eq.~\eqref{eq:ActionCollectiveFluctuations}, the integral extends over
    Goldstone mode fluctuations, $Q=T \tau_3 T^{-1}$. More precisely, we integrate over the
    invariant measure $DQ$
    of the Goldstone mode manifold, AIII$_{(2|2)}$, where 
    \begin{align}
        \label{eq:AIIIIntegrationManifold}
            \textrm{AIII}_{(n|n)}\equiv \textrm{U}(n|n)/(\textrm{U}(n/2|n/2) \times \textrm{U}(n/2|n/2)),
    \end{align}
     i.e. the coset space of symmetry transformations $T$ in
    $Q=T\tau_3 T^{-1}$, with transformations $[T,\tau_3]$ modded out. 
    The
    essential difference to the SYK $\sigma$-model is the absence of the
    collective mode fluctuations, reflecting the rigid spectral
    density of the matrix ensemble.

    The theory above  is set up to generate determinant ratios, 
    \begin{align}
        \label{eq:SigmaDeterminantRatio}
        Z(\hat \epsilon)\equiv \int dQ \, \exp \left( -S[Q] \right)=
        \left\langle\frac{\det(\epsilon^{+ \textrm{b}}-H)\det(\epsilon^{- \textrm{b}}-H)}{\det(\epsilon^{+ \textrm{f}}-H)\det(\epsilon^{- \textrm{f}}-H)}\right\rangle_H,
    \end{align} 
    from where observables such as the two-point function
    Eq.~\eqref{eq:MatrixCorrelationFunction} are obtained by differentiation: 
    \begin{align}
        \label{eq:MatrixSigmaModelCorrelationFunction}
        R_2(\omega)= \partial^2_{\epsilon^{+\textrm{b}},\epsilon^{-\textrm{b}}} Z(\hat \epsilon) \big|_{\hat \epsilon^{s\textrm{b}}=\hat\epsilon^{s\textrm{f}}=E+s \frac{\omega}{2}  }.
    \end{align}
    Note that the action in Eq.~\eqref{eq:ActionSigmaEpsilon} is proportional
    only to differences between the energy arguments (because $\str(Q E)=0$),
    i.e., it is proportional to the parameter $s=\pi \omega \rho$, indicating that the ladder
    diagram structure characterizing the  singular diagrams is hardwired into this
    description. To see how this materializes in concrete terms, consider the
    case $s>1$ (i.e. the regime of perturbation theory) where the largeness of
    the action requires $Q\approx \tau_3$, and $T=\exp(W)$ must be close to the identity. With a
    parameterization of the coset space in terms of the generators $W=\left( \begin{smallmatrix}
        & B \cr \tilde B
    \end{smallmatrix}
     \right) $ (off-diagonal in advanced/retarded space), we can expand the
     action in powers of $B,\tilde B$. For example, it is straightforward to
     verify that the expansion of the action to leading quadratic order reads
     $S^{(2)}[B]=-i s \,\mathrm{str}(B \tilde B) $ indicating that the
     `propagator', $\sim s^{-1}$ of the $B$-modes is the random matrix ladder
     diagram. At next leading order in the expansion we obtain
     $S^{(4)}[B]=-\frac{2is}{3}\mathrm{str}(B \tilde B B \tilde B)$, i.e. a
     structure with four $B$-ports, weighted by a factor $s$ --- the dark shaded
     cross. Expanding the pre-exponential factors in the same manner, the
     $B$-integral yields the result of the diagrammatic resummation in a
     comparatively painless manner. 
     
     Conceptually, we can think of the $Q(B)$-theory as the \keyword{effective
     zero-dimensional field theory} of quantum chaos. In this language, the
     topological contents of its diagrams is encoded in the loop structure of
     the $Q$-integral, which in turn reflects the geometry of the Goldstone mode
     manifold. However, the merit of this representation goes  beyond that of
     an efficient reproduction of diagrammatic results: First, the cancellation
     of all diagrams at order $g>0$ --- which in physical terms reflects the
     linearity of the GUE form factor in time --- is a manifestation of a
     peculiarity of the class A nonlinear $\sigma$-model, namely its
     \keyword{semiclassical exactness}~\cite{Zirnbauer:1996zz,Efetov:1983xg}: one can show that the integral
     Eq.~\eqref{eq:MatrixSigmaModelCorrelationFunction} is exactly given by the
     saddle point approximation around its two stationary points $\pm \tau_3$, i.e.
     all contributions beyond second order in $B$ in the expansion around these
     saddle points cancel out. Here, $\tau_3$ is the `standard' saddle point,
     whose fluctuation integral yields the ramp in the form factor, while the
     $-\tau_3$ \keyword{Altshuler-Andreev} saddle point~\cite{Andreev:1995qgf} is responsible for the
     plateau. 
     
     More generally, i.e. for matrix ensembles of different symmetry, one
     obtains $\sigma$-models of identical structure but with different
     integration manifolds. In these cases, the $B$-expansion need not
     terminate, leading, e.g., to the nontrivial form factor of the AI (GOE)
     model. Still the integrals can be carried out for arbitrary $s$, providing
     access to the full non-perturbative contents of the theory. The same
     information can of course be obtained from the powerful non-perturbative
     tools of random matrix theory. However, the advantage of the $\sigma$-model
     in the present context is that its scope extends beyond the random  matrix
     regime. We have already exemplified this in the context of the SYK model. In
     Section~\ref{sec:NonPerturbativeJT}, we will demonstrate the emergence of
     the $\sigma$-model within in the string theoretical completion
     of JT gravity, providing a (doubly non-perturbative) realization of the
     holographic program. 
\end{BoxTypeA}

\subsection{The spectral edge}\label{sec:spectralEdge}

\noindent\textit{In this section, the energy parameter $E$ is measured relative to
the gap edge.}\vspace{0.1cm}

Previously, we had mentioned that the holographic correspondence unfolds for
 energies close to the ground states of the participating theories. We also
 noted that the SYK model with its collective fluctuations supports a
 statistically distributed `edge' whose fluctuations with variance
 $\textrm{var}(E_0)\propto N^2$ exceeding the many body level spacing $\propto
 2^{-N/2}$ by far. 

\begin{figure}
    \centering
\includegraphics[width=0.7\textwidth]{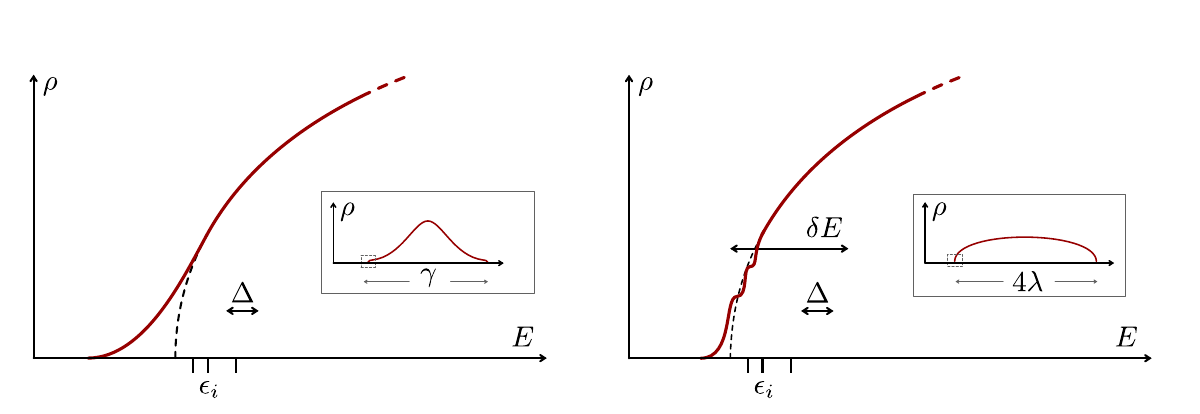}
\caption{Left: Schematic representation of the  spectral edge of the SYK model
with its broadly extended tail. Right: the more rigid spectral edge of an
invariant random matrix ensemble with superimposed fluctuations indicating near
crystalline order of last eigenvalues of the spectrum. }
\label{fig:SpectralEdge}
\end{figure}

Unitarily invariant  random matrix ensembles show no such fluctuations. Here,
the edge is pinned with near level spacing precision, see
Fig.~\ref{fig:SpectralEdge}, right. A closer look shows that the $\sim \sqrt{ E}$ of the $N\to
\infty$ `double scaled' spectral density is modified by an small ripples
separated by the near edge level spacing $\Delta_0\sim \lambda/D^{2/3}$ reflecting an almost crystalline order
of the levels above the edge. Quantitatively, these features are captured by the
Airy formula~\cite{Tracy:1992kc}
\begin{align}
    \label{eq:SpectralDensityEdgeDense}
    \langle \rho( E) \rangle = \frac{1}{\Delta_0} \left(\tilde  E\textrm{Ai}^2(-\tilde  E)+(\textrm{Ai}'(-\tilde  E))^2\right),\qquad \tilde  E\equiv E/\Delta_0.
\end{align}  
with this information, we can formulate holographic consistency checks of
increasing levels of ambition: 
\begin{itemize}
    \item One-point function in the double scaled limit: does the gravitational
    bulk match the $\sinh$-spectral density of the SYK model,
    Eq.~\eqref{eq:SYKDoS}?
    \item One-point function in the extreme edge vicinity: does the bulk show a
    `hard edge', as in random matrix theory  Eq.~\eqref{eq:SpectralDensityEdgeDense}?\footnote{Confusingly, RMT refers to 
    Eq.~\eqref{eq:SpectralDensityEdgeDense} as a `soft edge'. However, its
    softness manifests itself on level spacing scales, for our purposes, this is
    as `hard' as it gets.} This is  asking if the bulk is a dense or a sparse
    quantum chaotic system. 
    \item If the former is the case, can we access the `doubly
    non-perturbative' ripples in the near edge spectral density? (This would
    realize a gold standard: access to individual quantum states in a
    gravitational system!)
    \item Spectral correlations: in the double scaled limit, can the spectral
    correlations of the boundary theory, or the matrix theory proxy, be obtained
    from the gravitational theory?
\end{itemize}
In the rest of this review we will address all of these questions in turn.
\begin{BoxTypeA}[box:kontsevich]{Kontsevich model}

    Previously, in Box~\ref{box:sigma-model}, we had introduced the nonlinear
    $\sigma$-model as a tool to describe spectral correlations of matrix
    ensembles. Conceptually, the $\sigma$-model is stabilized by the
    double-scaling limit, where a large density of states serves as an order
    parameter of the causal symmetry breaking transition, with superimposed
    $Q$-matrix Goldstone model fluctuations. In the extreme vicinity of the
    edge, this symmetry breaking scenario is not yet developed, and the exact
    $\mathrm{U}(2|2) $ symmetry of the model remains largely unbroken. The
    \keyword{Kontsevich model } describes this regime in terms of the action
    \begin{align}
    \label{eq:KontsevichAction}
    S[A,\hat \epsilon]=\Delta_0^{-3/2}\,
\left(\str(A\hat E)+\frac{1}{3} \str(A^3)\right),
\end{align}
where $A$ is a $4\times 4$ supermatrix, and $\hat E=\mathrm{diag}(E_1^+,E_2^+,E_1^-,E_2^-)$
a matrix of energy arguments that can be chosen to compute the average spectral
density, or its correlations. (To compute $\langle  \rho(E) \rangle$ a reduced
$2\times 2$ matrix  suffices.) The linear+cubic structure of the action
implies that it is not suitable for a perturbative approach. However, the
reminiscence to integral representations of the Airy function indicates that the
result Eq.~\eqref{eq:SpectralDensityEdgeDense} is around the corner. Indeed, one
may obtain this formula by a straightforward integration over the flat measure
of $A$-matrices. However, the utility of the model goes beyond this:

The full integration over $A$-matrices over the integration domain
$\textrm{U}(2|2)$ describes the spectral structure of the model including in
regions with vanishing spectral density $E<0$. Physically, this is the ending of
the specral curve's  branch cut, or the  termination of the `causal symmetry broken' domain of quantum chaos.
Conversely, for energies 
$E/\Delta_0 \gtrsim 1$ inside the gap, the integral may be collapsed to the manifold of stationary
configuration of the $A$-matrices. A variation of the action 
$\delta_A S[A]=0$, shows that the cubic $A$-action leads to a quadratic stationary
equation. Considered as functions of the complex energy arguments $E^\pm$, its
two solutions define a double cover of the complex plane, in analogy to the
spectral curve, and establishing causal symmetry breaking. The $\sigma$-model
with action Eq.~\eqref{eq:ActionSigmaEpsilon}
then emerges as the Goldstone mode theory on top of these symmetry broken saddle
points, as an integral over stationary configurations, $A\sim i \rho(E) T \tau_3
T^{-1}$. 

More generally, the integration over `massive' fluctuations around the saddle
points yields corrections to the double scaled  spectral density $\langle
\rho(E) \rangle\sim \sqrt{E}$ in terms of ribbon-like diagrams. The first of
these corrections scales as $\delta \langle \rho(E) \rangle\sim
E^{-5/2}$~\cite{altlandQuantumChaosEdge2024}. This correction is a prediction of
random matrix theory at the spectral edge, and notably lies beyond the reach of
the JT disk path integral, which captures only the leading $\sinh\left( 2\pi
\sqrt{E}\right)$ density. What, then, is its gravitational origin? As we will
see in Section~\ref{sec:JTTopologicalRecursion}, the answer is topology: the
$E^{-5/2}$ scaling is precisely reproduced by the so-called handle-disk ($g=1$,
$n=1$) contribution to the topological
expansion~\cite{Altland:2025SpectralEdge}, providing a striking precision test
of the bulk--boundary correspondence beyond the leading saddle.  An even more ambitious
goal would be to identify the Kontsevich model itself as an effective theory in
a gravitational setting. We will return this point in Section
\ref{sec:NonPerturbativeJT}.
\end{BoxTypeA}

\subsection{Physics at level spacing scales}

Let us conclude this section with a few remarks on the description of chaotic
spectra at the highest resolution scale, i.e. the level spacing $\Delta$. Use
cases include the  level repulsion in spectral correlation
functions at scales $\lesssim \Delta$, their manifestation as a `plateau' in the spectral form factor at
time scales exceeding the Heisenberg time $2\pi/\Delta$, or the fine structure
of the spectral edge as described by the oscillatory Airy functions,
Eq.~\eqref{eq:SpectralDensityEdgeDense}. A defining feature of this regime is 
that it is not describable in terms of semiclassical `diagram' summations.
Conversely, theories that assume the form of series by design --- in the next
section, we will discuss how JT gravity falls into this category --- are
incapable of accessing it, or at least require support by some non-perturbative
completion scheme. 

This breakdown of perturbation theory can be understood from different
perspectives. One starts with the observation that the `causal symmetry
breaking' instrumental to the formulation of the nonlinear $\sigma$-model showed
in the formation of a large imaginary part $\epsilon \pm i 0 \to \epsilon \pm i
\gamma$, within a stationary phase approach treating the spectral density as a
continuous function (or, equivalently, the spectral resolvent as a quantity
containing branch cut described via the spectral curve.) However, the spectrum
of individual systems actually contains discrete poles whose positions need to
be individually resolved in the regimes mentioned above. In other words, they
are regimes of `unbroken symmetry', where the un-breaking proceeds via
non-perturbatively strong Goldstone mode fluctuations. Another way to arrive at
the same conclusion is to observe that our $\sigma$-models are `zero-dimensional'
theories, where continuous symmetry breaking does exist in the limit of
vanishing explicit symmetry breaking $\epsilon/\Delta \to 0$. 

In some cases, the effect of symmetry restauration can be captured by an
extended stationary phase analysis including (`Altshuler-Andreev') saddle points
perturbatively inaccessible from the causal ones. However, more generally, one
needs to integrate over the full Goldstone mode manifolds. In Section \ref{sec:NonPerturbativeJT} we will
discuss how these structures materialize in the gravitational context. 
\section{The topological expansion of the JT path integral }\label{sec:TopoJT}
\begin{figure}[t]
\centering
\includegraphics[width=0.5\textwidth]{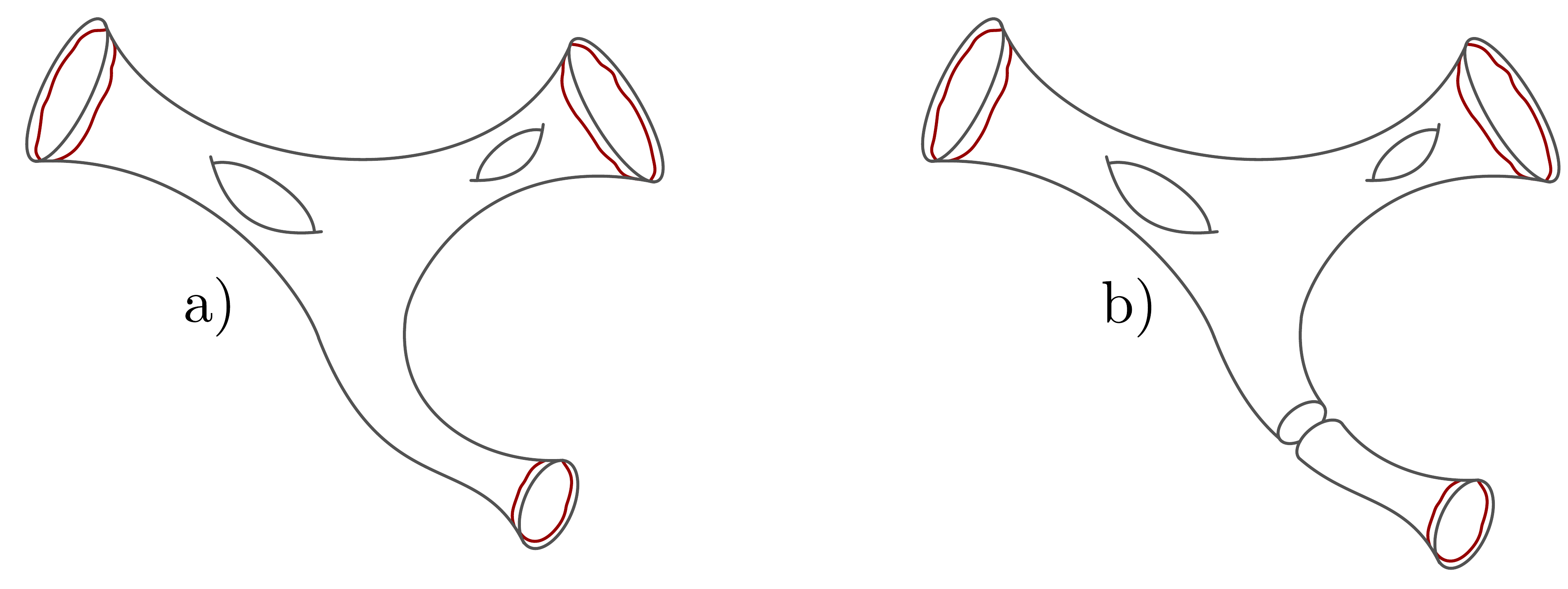}
\caption{a) An AdS spacetime with $n=3$ boundaries and $g=2$ internal handles. b) one asymptotic boundary has been separated along a geodesic, creating a geodesic boundary on the internal Riemann manifold, and leaving a trumpet geometry.}
\label{fig:JTMultipleBoundary}
\end{figure}

Having described the classical theory in Section \ref{sec:ClassicalJT} and the
necessary matrix theory background in Section~\ref{sec:MatrixTheory}, let us now
move on to the full quantum theory, described via the path integral
\begin{equation}\label{eq:JTPathIntegral}
 \boxed{{\cal Z}(\left\{\beta_i \right\}) =  \sum_{\rm top} \int \frac{{\cal D}[\Phi,g]}{{\rm Diff}} e^{-S_{\rm E}[\Phi,g]}}
\end{equation}
Implicit in this equation are several non-trivial choices and instructions that
 we will make more explicit in what follows. Firstly, as indicated, we sum over
 topologies, which in the context of Euclidean 2D gravity means a sum over
 Riemann surfaces of increasing genus $g$, at given fixed number of boundaries,
 $n$, cf. Fig.~\ref{fig:JTMultipleBoundary}. Secondly accounting for the local
 invariance of the gravity action, we mod out by Diff, the group of
 diffeomorphisms. This is effectively the gauge symmetry of JT gravity, as two
 metric and dilaton configurations are equivalent if they are diffeomorphic. The
 precise treatment of this diffeomorphism redundancy will play an important role
 in the discussion that follows.  Thirdly, the argument of the partition
 function indicates that the Euclidean geometries we sum over have specified
 boundary conditions, consisting of a disjoint union of circles, the
 circumference of each is given by the corresponding $\beta_i$ in the list of
 arguments, with the dilaton boundary condition fixed to $\Phi_{r,i}/\epsilon$
 at each boundary component labeled by $i$. Lastly, the exponent $S_{\rm
 E}[\Phi,g]$ is given by the Euclidean action of JT gravity,
\begin{equation}\label{eq:EuclideanJT}
S_E
=
-\frac{\Phi_0}{2}
\left(
\int_{\mathcal M}\sqrt{g}\,R
+2\int_{\partial\mathcal M}\sqrt{\gamma}\,K
\right)
-\frac{1}{2}
\int_{\mathcal M}\sqrt{g}\,\phi\,(R+2)
-
\int_{\partial\mathcal M}\sqrt{\gamma}\,\phi\,(K-1),
\end{equation}
where $\Phi = \Phi_0 + \phi$, giving rise to the topological term coupling
$\Phi_0$ to the Euler character, $\chi(M)$, introduced in
\eqref{eq:EulerCharacter}, and the fluctuating part of the dilaton, $\phi$.
 It will be useful to
remember that the first two terms are topological, and that the last term is a
pure boundary integral. Geometrically, we deal with spacetimes of the form shown
in Fig.~\ref{fig:JTMultipleBoundary} a), which have $n$ asymptotic AdS$_2$ type
boundaries and $g$ internal handles. 

The path integral \eqref{eq:JTPathIntegral} is over both the dilaton and the
metric. The dilaton can be dealt with easily, paralleling the situation for the
classical theory. In the quantum theory, integrating over $\phi$ simply imposes
the constraint $\delta(R+2)$ inside the path integral, which therefore now
ranges over hyperbolic metrics only. With the dilaton constraint kept explicit,
we are thus left with the expression
\begin{equation}\label{eq:ActionWithDilatonConstraint}
     {\cal Z}(\left\{\beta_i \right\}) = \sum_{\text{topologies}}e^{S_0 \chi}
\frac{1}{\mathrm{Vol}(\mathrm{Diff})}
\int \mathcal{D}g\;
\delta(R+2)\;
\exp\!\left[
\int_{\partial\mathcal M}
\sqrt{\gamma}\,\phi\,(K-1)
\right] = \sum_{g}e^{S_0 \chi} {\cal Z}_{g,n} (\{ \beta_i \}).
\end{equation}
As we saw, the topological coupling $e^{S_0\chi}$, with $S_0 = 2\pi\Phi_0$, arises from the first two terms in \eqref{eq:EuclideanJT} via the Gauss--Bonnet theorem \eqref{eq:EulerCharacter}. 

As it stands, the gravitational path integral  still contains too much
structure for a direct comparison to matrix theory. At the same time, it extends over a large class of geometries, solely constrained to have
constant negative curvature --- the $\delta(R+2)$ factor. This implies a huge
level of symmetry which in turn allows a reduction of the integral to a much
simpler sum over topologies, weighted by volume factors, the so-called
Weil-Petersson volumes. Intuitively, this reduction can be described in
terms of a surgical procedure:  imagine `chopping off' each asymptotic boundary
along a geodesic boundary of length $L_i$, leaving a Schwarzian trumpet as
described in Section \ref{sec:trumpetIntegral}. What then remains is the set of
trumpets with geodesic boundaries of lengths $\{ L_i \}$, as well as an internal
Riemann surface $\Sigma_{g,n}$ of genus $g$ with the same set of geodesic boundaries, cf.
Fig.~\ref{fig:JTMultipleBoundary} b). The path integral itself mathematically
decomposes into the same constituents, that is $n$ Schwarzian trumpet
contributions $Z_{\rm tr} (\beta_i, L_i)$, as well as an integral over the
remaining Riemann surface of genus $g$ with $n$ geodesic boundaries,
$\Sigma_{g,n}$, which results in the \textbf{Weil-Petersson volume} $V_{g,n}(\{
L_i\})$. The final path integral, for each topology, is then given by the gluing
formula
\begin{equation}\label{eq:TrumpetGluingTopology}
\boxed{{\cal Z}_{g,n} (\{ \beta_i \}) = \int_0^\infty
\prod_{i=1}^n \left(L_i\, dL_i\right) Z_{\mathrm{tr}}(\beta_i,L_i)V_{g,n}(\{ L_i \})}
\end{equation}
The full partition function is the sum of the ${\cal Z}_{g,n}$ over
genus $g$, and in Section \ref{sec:JTTopologicalRecursion} we will discuss it as
starting point of a geometrically formulated topological recursion procedure
quantitatively equivalent to that of matrix theory. Readers primarily interested
in the backbone of the holographic construction  may jump directly to that
section. For those ready to dwell a little more on its concrete realization we
summarize the, exquisitely beautiful, mathematics of the reduction of
Eq.~\eqref{eq:ActionWithDilatonConstraint} to
Eq.~\eqref{eq:TrumpetGluingTopology} in Section \ref{sec:PathIntegralReduction} below. 

\begin{BoxTypeA}[box:wp-volumes]{Weil-Petersson volumes}

The geometry of the Weil-Petersson volumes featuring in
Eq.~\eqref{eq:TrumpetGluingTopology} is intuitive and can be described without
invoking the full apparatus underlying their construction:   instead of one unique Riemann surface
$\Sigma_{g,n}$ of genus $g$ with $n$
geodesic boundaries of lengths $L_1, \ldots, L_n$, there is a whole family of
such geometries, all connected by continuously deforming the shape without
changing topology. This family of continuous deformations is itself parametrised
by \keyword{moduli space} $\mathcal{M}_{g,n}$. The total volume of this moduli
space is the \textbf{Weil-Petersson
volume} 
\begin{equation}
    V_{g,n}(L_1, \ldots, L_n) = \int_{\mathcal{M}_{g,n}} d\mu_{\rm WP}\,,
\end{equation}
where canonical measure $d\mu_{\rm WP}$ arises from a  symplectic
structure on $\mathcal{M}_{g,n}$. These volumes appear in the path integral of
JT gravity as part of the gluing construction described in this section. If we
start from a given contribution to the path integral with $n$ asymptotic
boundaries and then cut off the asymptotic boundaries along internal geodesics
of length $\{ L_i \}$,  we are left with the `bulk' internal Riemann surface of
genus $g$ with $n$ geodesic boundaries. Given that the path integral runs over
{\it all possible} such Riemann surfaces, and the gauge-fixed gravitational
measure is precisely the Weil-Petersson measure, the Riemann surface contributes
a factor $V_{g,n}$ to the path integral. This volume thus counts the all
possible interior geometries of given topology, weighted by the JT gravity
measure.

To build intuition for the moduli that are being integrated over, it helps to
describe them in terms of the so-called Fenchel-Nielsen coordinates, $(\tau_i,
\ell_i)$, which constitute the canonical set of coordinates pertaining to the
symplectic structure mentioned above. Any Riemann surface $\Sigma_{g,n}$ can be
cut along $(3g - 3 + n)$ internal geodesics into pairs of pants (three-holed
spheres). The moduli are then the lengths $\ell_i$ and twist angles $\tau_i$ of
these internal geodesic cuts, and the Weil-Petersson measure takes the simple
form $d\mu_{\rm WP} = \prod_{i=1}^{3g-3+n} d\ell_i\, d\tau_i$. The simplest
case, $V_{0,3}(L_1, L_2, L_3) = 1$, says that a pair of pants with three fixed
boundary lengths has a unique hyperbolic metric --- there are no remaining
moduli. The crucial property that makes the topological expansion
computationally tractable is that the $V_{g,n}$ satisfy Mirzakhani's recursion
relation, which expresses $V_{g,n}$ in terms of volumes of simpler surfaces
obtained by cutting along a geodesic, relating the original surface to either a
lower-genus surface or a pair of simpler components.
\end{BoxTypeA}

\subsection{From Eq.~\eqref{eq:ActionWithDilatonConstraint} to
Eq.~\eqref{eq:TrumpetGluingTopology}}
\label{sec:PathIntegralReduction}

The key to the simplifcation of the gravitational path integral lies in its 
diffeomorphism invariance: two hyperbolic metrics $g$ and $\varphi^*g$ related by a
diffeomorphism $\varphi$ describe the same physical geometry, and the naive path
integral $\int {\cal D}g$ overcounts by the (infinite) volume of the
diffeomorphism group. Just as in Yang-Mills theory, a convenient
tool\footnote{Note that in the JT case other approaches also exist, and are
arguably more elegant, such as the first-order formalism utilised in
\cite{SaadShenkerStanford2019JTMatrixIntegral}. However, these necessitate more
background material and are  beyond the scope of this review.} for handling
this is the Fadeev-Popov procedure. In the following, we first describe the space we want to end
up on, and then explain how Fadeev-Popov gets us there.

\paragraph*{Teichm\"uller space and moduli space } The physical configurations, hyperbolic metrics modulo diffeomorphisms, form the \keyword{moduli space}
\begin{equation}\label{eq:ModuliSpaceDef}
    {\cal M}_{g,n} = \bigl\{\textrm{hyperbolic metrics on } \Sigma_{g,n}\bigr\}\big/\mathrm{Diff}\,.
\end{equation}
To describe this space more concretely, it helps to note that the diffeomorphism
group has two qualitatively different components. First, there are small
diffeomorphisms, continuously connected to the identity, forming the subgroup
$\mathrm{Diff}_0$. Second, there are \textit{large} diffeomorphisms ---
topologically non-trivial rearrangements that cannot be continuously shrunk to
the identity, such as cutting the surface along a closed geodesic, rotating one
side by $2\pi$, and re-gluing. These form the discrete \textbf{mapping class
group},
\begin{equation*}
    \mathrm{MCG}_{g,n} = \mathrm{Diff}\,/\,\mathrm{Diff}_0\,.
\end{equation*}
It is therefore natural to perform the quotient \eqref{eq:ModuliSpaceDef} in two
stages. Quotienting first by $\mathrm{Diff}_0$ alone gives \textbf{Teichm\"uller
space},
\begin{equation}\label{eq:TeichmuellerDef}
    {\cal T}_{g,n} = \bigl\{\textrm{hyperbolic metrics on } \Sigma_{g,n}\bigr\}\big/\mathrm{Diff}_0\,.
\end{equation}
This is a smooth, simply connected manifold of real dimension
\begin{equation}\label{eq:TeichDim}
    \dim_{\mathbb{R}}\, {\cal T}_{g,n} = 6g - 6 + 2n\,.
\end{equation}
Each point in ${\cal T}_{g,n}$ represents a genuinely distinct hyperbolic shape: all the continuous redundancy has been removed, and what remains is a finite-dimensional space of physical deformations. Moduli space is then the further quotient by the discrete MCG,
\begin{equation}\label{eq:ModuliTeichMCG}
    {\cal M}_{g,n} = {\cal T}_{g,n}\,/\,\mathrm{MCG}_{g,n}\,.
\end{equation}
This two-stage structure is important because Fadeev-Popov and the MCG identification play different roles, as we now explain.

\paragraph*{Fadeev-Popov gauge fixing}

The Fadeev-Popov procedure handles the continuous part of the gauge group,
$\mathrm{Diff}_0$. In standard fashion, we insert
\begin{equation}\label{eq:FPInsertion}
    \int \frac{{\cal D}g}{\mathrm{Vol}(\mathrm{Diff}_0)} = \int {\cal D}g\;\Delta_{\mathrm{FP}}[g]\;\delta(F[g])\,,
\end{equation}
where $F[g]=0$ is a gauge-fixing condition and $\Delta_{\mathrm{FP}}$ the
associated Fadeev-Popov determinant. The key step is to decompose an
infinitesimal metric variation into a pure diffeomorphism and a genuine moduli
deformation,
\begin{equation}\label{eq:MetricDecomposition}
    \delta g_{\mu\nu} = \underbrace{\nabla_\mu \xi_\nu + \nabla_\nu \xi_\mu}_{\text{gauge (diffeo)}} + \underbrace{\sum_I \delta m^I \,\mu^{(I)}_{\mu\nu}}_{\text{physical (moduli)}}\,.
\end{equation}
The first term, $\nabla_\mu \xi_\nu + \nabla_\nu \xi_\mu$, is a Lie derivative
along the vector field $\xi^\mu$, describing how the metric changes under an
infinitesimal coordinate transformation, and is therefore pure gauge. The second
term captures the \textit{physical} metric deformations: those that genuinely
change the shape of the surface. These are parameterised by the moduli $m^I$, $I
= 1, \ldots, 6g - 6 + 2n$, and the tensors $\mu^{(I)}_{\mu\nu}$ (known as
Beltrami differentials) are simply a basis for these physical deformations; they
span the directions in metric space that are transverse to the gauge orbits and
thus correspond to tangent vectors of Teichm\"uller space at the given metric.
This decomposition defines a change of variables
\begin{equation}
    {\cal D}g_{\mu\nu}\quad \longrightarrow \quad {\cal D}\xi^\mu \times \prod_I dm^I\,.
\end{equation}
The integral over ${\cal D}\xi^\mu$ cancels $\mathrm{Vol}(\mathrm{Diff}_0)$, and
the Jacobian of the change of variables is precisely the Fadeev-Popov
determinant $\Delta_{\mathrm{FP}}$. What survives is an integral over the
finite-dimensional moduli $m^I$, $I = 1, \ldots, 6g-6+2n$, with a measure given
by $\Delta_{\mathrm{FP}}$ evaluated on the gauge slice.

\paragraph*{From Fadeev-Popov to the Weil-Petersson measure}

We now need to settle the question what measure on Teichm\"uller space the
Fadeev-Popov determinant actually produces. To answer this, we use the fact that
${\cal T}_{g,n}$ admits a natural set of global coordinates, the so-called \textbf{Fenchel-Nielsen coordinates}. These are constructed by decomposing
$\Sigma_{g,n}$ into pairs of pants (three-holed spheres) by cutting along $(3g -
3 + n)$ internal geodesics. Each cut $i$ is characterised by two parameters:
\begin{itemize}
    \item the geodesic length $\ell_i > 0$ of the cut, and
    \item a twist angle $\tau_i \in \mathbb{R}$, specifying how the two adjacent
    pants are glued along that geodesic.
\end{itemize}
The $2 \times (3g - 3 + n)$ parameters $(\ell_i, \tau_i)$ provide global
coordinates on ${\cal T}_{g,n}$, consistent with the dimension stated in
\eqref{eq:TeichDim}. In these coordinates, the Fadeev-Popov determinant takes a
remarkably simple form: it combines with the ultralocal norm on the space of
metrics to yield the measure
\begin{equation}\label{eq:WPMeasure}
    d\mu_{\mathrm{WP}} = \prod_{i=1}^{3g-3+n} d\ell_i\, d\tau_i\,.
\end{equation}
This is the \textbf{Weil-Petersson measure}, and the fact that Fadeev-Popov
produces it is not accidental. A result due to Wolpert~\cite{Wolpert1983} shows that the
Weil-Petersson symplectic form on Teichm\"uller space is
\begin{equation}\label{eq:WPSymplecticForm}
    \omega_{\mathrm{WP}} = \sum_{i=1}^{3g-3+n} d\ell_i \wedge d\tau_i\,,
\end{equation}
meaning that the lengths and twists forming the Fenchel-Nielsen coordinates are canonically conjugate variables, just
like positions and momenta. The measure \eqref{eq:WPMeasure} is the canonical
volume form $d\mu_{\mathrm{WP}} = \omega_{\mathrm{WP}}^{3g-3+n}/(3g-3+n)!$
determined by this symplectic structure. In other words, the Fadeev-Popov
procedure applied to 2D gravity produces the unique measure dictated by the
symplectic geometry of Teichm\"uller space. (This can also be understood without
gauge fixing, directly from the symplectic structure; see e.g.~Wolpert 1985.)

\subsubsection*{The gluing formula}

Assembling all ingredients, the gauge-fixed JT path integral on a genus-$g$, $n$-boundary topology reads
\begin{equation}\label{eq:GluingFormulaDerived}
   \boxed{ {\cal Z}_{g,n}(\{\beta_i\}) = \int_{{\cal M}_{g,n}} d\mu_{\mathrm{WP}}\;\prod_{i=1}^n \int_0^\infty L_i\, dL_i\; Z_{\mathrm{tr}}(\beta_i, L_i)}
\end{equation}
reproducing \eqref{eq:TrumpetGluingTopology} with
\begin{equation}
    V_{g,n}(L_1, \ldots, L_n) = \int_{{\cal M}_{g,n}} d\mu_{\mathrm{WP}}\,,
\end{equation}
confirming that the Weil-Petersson volumes are the gauge-fixed gravitational path integral over the interior Riemann surface. 

 As a
concrete illustration of this machinery, consider the first subleading
correction to the spectral density, which is obtained as the inverse Laplace
transform of the JT gravitational path integral with a single asymptotic
boundary of length $\beta$. Note that this gives the leading correction to the
disk path integral of the Schwarzian discussed in \eqref{sec:ClassicalJT}. We
thus look at the term corresponding to $g=1$, $n=1$ in the topological
expansion. The relevant Weil-Petersson volume is $V_{1,1}(L) = \frac{1}{48}(L^2
+ 4\pi^2)$. Substituting into the gluing
formula~\eqref{eq:TrumpetGluingTopology} together with the trumpet partition
function~\eqref{eq:Ztrumpet}, and performing the inverse Laplace transform, one
finds that the handle-disk contributes a correction $\delta\rho(E) \sim e^{-S_0}
E^{-5/2}$ to the spectral density~\cite{Altland:2025SpectralEdge}. This is
suppressed by one power of $e^{-S_0}$ relative to the disk, reflecting the Euler
characteristic $\chi = 2 - 2\cdot 1 - 1 = -1$. The $E^{-5/2}$ scaling precisely
matches the prediction from the Kontsevich model discussed in
Box~\ref{box:kontsevich}, providing a quantitative confirmation that the first
subleading correction to the Schwarzian density has a geometric origin in the
topology of the gravitational path integral.

\subsection{From the JT path integral to topological recursion}
\label{sec:JTTopologicalRecursion}
With the contribution of surfaces of definite genus and boundary number
given by Eq.~\eqref{eq:GluingFormulaDerived}, the full partition function is obtained by summation over these topological sectors, 
\begin{equation}\label{eq:TopoExpansion}
    {\cal Z}(\{\beta_i\}) = \sum_g e^{S_0(2-2g-n)}\, {\cal Z}_{g,n}(\{\beta_i\})\,,
\end{equation}
where we have written out the Euler characteristic $\chi = 2-2g-n$ for a Riemann surface of genus $g$ with $n$ boundaries.

\begin{figure}[h]
\centering
\includegraphics[width=1\textwidth]{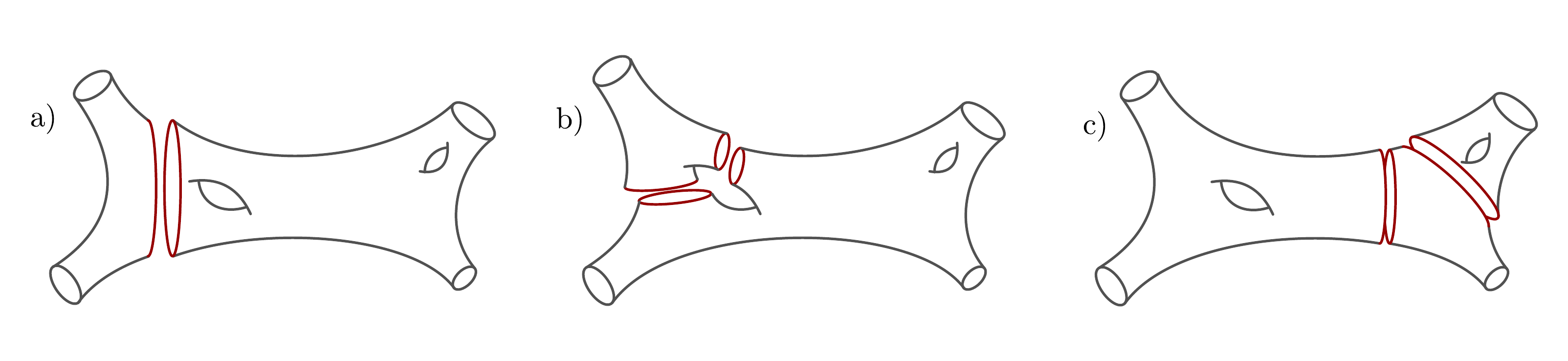}
\caption{Topological recursion in terms of pants decomposition. When removing a
 pair of pants from a given Riemann surfaces with boundaries, $\Sigma$, to
 produce the `stripped' surface $\Sigma'$, three things can happen. a) the
 stripped surfaces $\Sigma'$ has one boundary less than $\Sigma$ had, but the
 genus remains unchanged. b) $\Sigma'$ has one boundary more than $\Sigma$ had,
 and $\Sigma'$'s genus is reduced by one. c) The surface $\Sigma$ is split into
 two disjoint surfaces in such a way that the total genus and remaining
 boundaries are distributed among the remaining pieces.}
\label{fig:TopologicalRecursion}
\end{figure}

A remarkable fact about the topological expansion formula
\eqref{eq:TopoExpansion} is that it satisfies a recursion relation inherited
from the recursion relation of Weil-Petersson volumes discovered by Mirzakhani.
Intuitively this recursion comes from the fact that one may cut up a given
Riemann surface along cycles to produce Riemann surfaces of lower genus whose
moduli space integrals are less complex. The different ways of accomplishing
such a cutting produce relations is analogous to Schwinger-Dyson equations in
quantum field theories,  allowing us to recursively generate the $V_{g,n}$
starting with lowest-genus input only, which is the rigid base case of $V_{0,3}
(L_1,L_2,L_3)=1$. The corresponding geometric structure is shown in Figure
\ref{fig:TopologicalRecursion}. In fact, the bulk picture of these relations can
be derived from the Schwinger-Dyson equations of a so-called `Universe
Field Theory', \cite{PostvanDerHeijdenVerlinde2022UniverseFieldTheory}, as
reviewed in Section~\ref{sec:NonPerturbativeJT}. To apply
this to the JT topological expansion, as was done in \cite{Saad2019}, it is best
to focus on the Laplace transformed objects,
\begin{equation}
    W_{g,n}(z_1,\dots,z_n)=\int_0^\infty
\prod_{i=1}^n\left(dL_i\, e^{-z_i L_i}\right)V_{g,n}(L_1,\dots,L_n),
\end{equation}
which satisfy the recursion
\begin{equation}
    W_{g,n}(z_1,\dots,z_n)=\sum_{\text{branch points}}\mathrm{Res}_{z\to a}
K(z_1,z)\Big[W_{g-1,n+1}(z,\sigma(z),z_2,\dots,z_n)+\sum_{g_1+g_2=g} \sum_{I\sqcup J=\{2,\dots,n\}}W_{g_1,|I|+1}(z,z_I)W_{g_2,|J|+1}(\sigma(z),z_J)
\Big]\,,
\end{equation}
exactly mirroring the geometric structure in Figure
\ref{fig:TopologicalRecursion}. The first term in the angular brackets comes from
cases where a handle is cut, thereby reducing the genus by one and increasing
the number of boundaries by one. The remaining terms come from cuts that
separate the surface into two components of genus $g_1$ and $g_2$ respectively.
Once we have the $W_{g,n}$ we can inverse Laplace transform, and plug them
into the gluing formula \eqref{eq:TrumpetGluingTopology} to generate the terms
in the JT expansion.

A remarkable fact about the JT recursion is that, in the form of the
Laplace-transformed expansion relation of $W_{g,n}$, it coincides exactly with
the topological recursion expansion of a class of matrix integrals, originally
discovered by Eynard and Orantin \cite{eynardRandomMatrices2018}. This form of
topological recursion was first introduced for finite-size hermitian matrix
models and applied to the double-scaled JT matrix model in \cite{Saad2019}.

\begin{BoxTypeA}[box:jt-matrix-model]{Double-scaling and the JT matrix model}

The concept of double scaling a matrix integral has been evoked several times,
see for example the Box just above Section \ref{sec:singularCorrections}. 
  As discussed there, the near-edge level density takes on the universal square-root form
  $\rho(E) \sim \sqrt{E}$, while the near-edge level correlations are given by the Airy kernel,  
  \eqref{eq:SpectralDensityEdgeDense}. Such double-scaling limits also exist for more elaborate
  matrix models, obtained by tuning the coupling constants to (multi-)critical points, revealing 
  universal double-scaled critical theories~\cite{Douglas:1989ve,Brezin:1990rb,Gross:1989vs}. A celebrated series is that of the  
  $(2,2m{-}1)$ minimal models, with potentials $V(H) = \sum_{k=1}^{m} g_{2k} H^{2k}$, which are
  related to $c<1$ string theory~\cite{DiFrancesco:1993cyw}. For finite $m$, the double-scaled theory is governed
   by a finite-order ODE, the so-called string equation, here the $m$-th member of the Painlevé/KdV hierarchy,   
  whose solution, with appropriate boundary conditions, defines the exact non-perturbative  
  answer. JT gravity can be viewed as a limiting member of this family (formally $m \to \infty$).
   It admits an analogous completion via the full KdV string equation~\cite{Johnson:2019eik,Johnson:2020exp,Johnson:2020jza}, though the non-perturbative answer is no longer unique. An alternative route to a non-perturbative definition, which we discuss below, is provided by the explicit finite-$N$ matrix model of \cite{Jafferis:2022JTMatter}.                                              
                                               
  This matrix model is constructed so that its double-scaling limit reproduces the topological
  expansion of the SSS matrix model to all orders. The matrix potential is given as an expansion 
  in Chebyshev polynomials,
  \begin{equation}                                                V(x) = \sum_{n=1}^{\infty} \frac{(-1)^{n-1}}{n} q^{n^2/2} \Big( q^{n/2} + q^{-n/2} \Big)
  T_{2n}\left(\frac{x}{a}\right),
  \end{equation}
  where $q \in [0,1]$ is a deformation parameter and $a = 2/\sqrt{1-q}$ sets the spectral edge.  
  In the limit $q \to 1$, the edge $a$ diverges, so the limit cannot be taken at face value: one 
  must simultaneously double-scale, sending $N \to \infty$ while zooming in on the lower edge of
  the spectrum. This double-scaling limit yields the $\sinh(2\pi\sqrt{E})$ spectral density of JT
   gravity, together with the correct level correlations at all orders in the genus expansion.

\end{BoxTypeA}

 In conclusion, we thus have a chain of connections
\begin{equation}
    \boxed{\text{JT gravity} \longrightarrow V_{g,n} \rightarrow \text{Mirzakhani recursion} \longrightarrow \text{JT topological recursion} 
    \longrightarrow \text{Eynard-Orantin recursion} \longrightarrow \text{RMT}}
\end{equation}
This sequence can also be read from right two left, making it a true
duality relation. Notice, however, that the construction
is perturbative in the parameter $e^{-S_0}$, and therefore does not yet
establish the full correspondence between gravity and a matrix theory. In fact,
one may sharpen this point  by noting that JT gravity is a series in
$\exp(-S_0)$, realized as topological sectors, by design. In this regard, it
resembles a `semiclassical proxy' of a genuine quantum theory, in line with the
interpretation of gravitational actions as semiclassical limits. We will return
to this point in the next section, where we will discuss the non-perturbative
completion of JT gravity.

\subsection{An ensemble dual for a single gravitational bulk?}
Before delving further into the topological expansion of JT gravity, let us
pause and discuss one of the most surprising implications of our results up to
here. From the point of view of bulk JT gravity, we are working with a {\rm
single} bulk theory, analogous for example to the higher-dimensional type IIB
string theory in the original Maldacena duality. Nevertheless, allowing for
non-trivial topology, we find that there are non-zero connected contributions to
multi-boundary correlators, the simplest being the two-partition function
correlator
\begin{equation}\label{eq:connectedPartitionFunction}
   \overline{ {\cal Z}(\beta_1) {\cal Z}(\beta_2)}\Bigr|^{\rm JT}_{\rm c} =  \left\langle {\cal Z}_{n=2}(\beta_1,\beta_2)\right\rangle_{\rm matrix}\,,
\end{equation}
which as discussed above, is perturbatively equal to the free energy of
the SSS matrix model on a two-boundary topology. The LHS of the equation
denotes the result of the JT path integral on a geometry with $n=2$ two
asymptotic boundaries, while the RHS denotes the dual quantity evaluted in the
SSS matrix model, which is an average of \textit{two} partitions functions with
respect to the matrix model measure. From the point of view of the matrix model,
it is not surprising that two partition functions, and therefore via inverse
Laplace, also two spectral densities, should have a connected correlation, $\left
\langle \rho(E_1) \rho(E_2)\right \rangle_{\rm c}$; this is exactly the
mathematical expression of universal chaotic level correlations we discussed in
Section \ref{sec:MatrixTheory} above. From the point of view of bulk gravity, on
the other hand, this is nothing short of astonishing! The {\rm single} theory of
JT gravity seemily encodes {\it level correlations}, and moreover those of a
matrix model. This calls for further investigation and explanation. 

\section{Non-perturbative gravity and the discreteness of the spectrum}
\label{sec:NonPerturbativeJT}

 JT topological recursion establishes a correspondence to a matrix model with
near edge spectral density~\eqref{eq:SpectralDensityJT}. We also know that,
within matrix theory, near edge correlations are described by the Kontsevich
model Eq.~\eqref{eq:KontsevichAction}.\footnote{As written, the Kontsevich
action describes the model in the double-scaling limit, where
$\sinh(2\pi\sqrt{E})\sim 2\pi \sqrt{E}$. For a version of this model resolving
the $\sinh$-profile, see
Ref.~\cite{AltlandPostSonnerVanDerHeijdenVerlinde2023QuantumChaos2DGravity}. }
This begs the question whether there might be a direct link between bulk gravity
and the Kontsevich model. Answering it  is an ambitious goal which requires
going beyond perturbation theory; it necessitates a doubly non-peturbative
understanding of the bulk, that is some kind of    UV completion of JT. 

Our so far theory has been formultated in terms of  expansions  in ascending
orders in the  coupling $g_s = e^{-S_0}$. In fact, these series have been
asymptotic, as can be seen from the higher-genus free energies which scale as
${\cal Z}_{g,n} \sim (2g)!$. Typically, such factorially divergent asymptotic
series signal the presence of further non-perturbative sectors, that is
contributions of order $e^{-A/g_s} = e^{-A e^{S_0}}$. Written in terms of the JT
coupling $e^{-S_0}$ these effects are doubly non-perturbative in the entropy
factor $S_0$. This makes them especially interesting, but also challenging, from
the point of view of the gravity interpretation. 

It is here where concepts of
quantum chaos  have offered decisive  insight into the discrete realm of quantum
microstate correlations in gravity. From the former point of view  doubly
non-perturbative effects are  related to the physics of the plateau. An identification of the plateau within the bulk
theory would thus resolve gravitational microstates at the level of the discretuum. 
Both from the matrix-model, and from the gravitational
perspective one can approach this program from a number of different angles,
many of which have been explored in the literature.
\begin{itemize}
    \item Non-perturturbative completion of the matrix model. One can postulate that the above perturbative equivalence of JT and the SSS matrix
    model smakes the matrix model the `true quantum
    description' of gravity, including in the non-perturbative regime. The goal is then to
    seek a UV completion of the SSS matrix model, which may however not be
    unique. This approach was advocated and begun to be explored in the original
    paper \cite{Saad2019}, and pushed further in
    \cite{Johnson:2019eik,Johnson:2020exp,Johnson:2020jza} in analogy
    with the minimal string program.
    \item One may use the techniques and insights of the resurgence
    program~\cite{Marino:2015yie} directly on the large-genus results of the
    JT path integral and explore the non-perturbative completions in this
    framework~\cite{Gregori:2021tvs,Eynard:2023qdr,Schiappa:2023ned}. The
    effects so uncovered have a natural physical
    interpretation as eigenvalue tunneling events in the dual matrix
    model~\cite{Shenker:1990uf,David:1990sk}, equivalently ZZ-brane
    contributions in the gravity
    description~\cite{Alexandrov:2003nn}. Such an approach gives precise
    mathematical results, but
    the question of which non-perturbative completion is physically selected
    remains open.
    \item An approach somewhat in between the previous two, the so-called
    $\tau-$scaling limit attempts an all-genus resummation of the perturbative
    ramp physics, which is possible within the double scaling limit $t,e^{S_0}
    \rightarrow \infty$, with $\tau := t e^{-S_0}$ held fixed. In this limit the
     lowest non-trivial eigenvalue can be accessed,
    while the nature and uniqueness of the ultimate UV completion remain open.
    \item In what arguably constitutes the most ambitious approach, one seeks a
    direct  completion of JT gravity, extending it through branes and
    potentially other stringy types of degrees of freedom. This approach is pursued by
    \cite{PostvanDerHeijdenVerlinde2022UniverseFieldTheory,
    AltlandPostSonnerVanDerHeijdenVerlinde2023QuantumChaos2DGravity}, who
    propose and analyze the so-called Universe Field Theory (UFT) of JT gravity
    in the form of the Kodaira-Spencer theory. The latter conceptualizes JT
    gravity itself as a  reduction of a higher-dimensional parent, some of the  degrees
    of freedom integrated over in the process effectively realizing an ensemble.
    An attractive feature of this particular UV extension is that the emergence of statistical
    correlations and  their non-perturbative completion naturally condition
    each other. In the next section, we will discuss its realization in some
    more detail. 
\end{itemize}

\subsection{The Universe Field Theory of JT gravity}
\label{sec:UniverseFieldTheory}
In this section, we discuss a non-perturbative completion of JT gravity using technology borrowed from string theory.  
While we try to keep the technical level of detail to a minimum, some
familiarity with the vocabulary of string theory will be required (cf. Ref.~\cite{Tong:2009np} for a concise introduction). Even then, the completion may feel
exotic, as it makes heavy reference to concepts of \textbf{string field theory},
which is a `second quantised' version of string theory. At the same time, the
results  of the construction will be quite concrete, and we here summarize
them for the benefit of readers inclined to skip ahead to Section~\ref{sec:StringTheoryResults}: 
\begin{itemize}
    \item Following a textbook paradigm, we here consider semiclassical gravity
    as a limit of some string theory. Specifically,  two-dimensional JT gravity is realized by
    dimensional reduction from a string theory with a six-dimensional target
    space. 
    \item This reduction involves the integration over a large number of
    internal degrees of freedom, effectively introducting the ensemble average
    we found expressed through the JT path integral. 
    \item In concrete terms, it leads to a two-dimensional conformal field
    theory subject to a qubic non-linearity, whose perturbative expansion
    reproduces that of the JT path integral. 
    \item However, in limits where we probe observables in the ultra-long time
    limit (the plateau), perturbation theory breaks down. Instead, the theory
    reduces to the Kontsevich matrix model, and upon double-scaling to the
    nonlinear $\sigma$-model. 
    \item A nice feature of this construction is that it yields the nonlinear
    $\sigma$-model entirely in terms of elements of the string theory; no
    reference to an `extraneous' matrix ensemble is made. (Rather, data
    equivalent to that of a matrix ensemble is introduced through a large number
    of open string amplitudes which are integrated over in the process of
    dimensional reduction.) 
    \item In this way several boxes are ticked at once:  the description
    of the gravitational bulk down to the level of its microstates, explicit
    demonstration of chaos in a simple string theory, and ensemble
    average demystified. 
\end{itemize}

\begin{figure}[h]
\centering
\includegraphics[width=.5\textwidth]{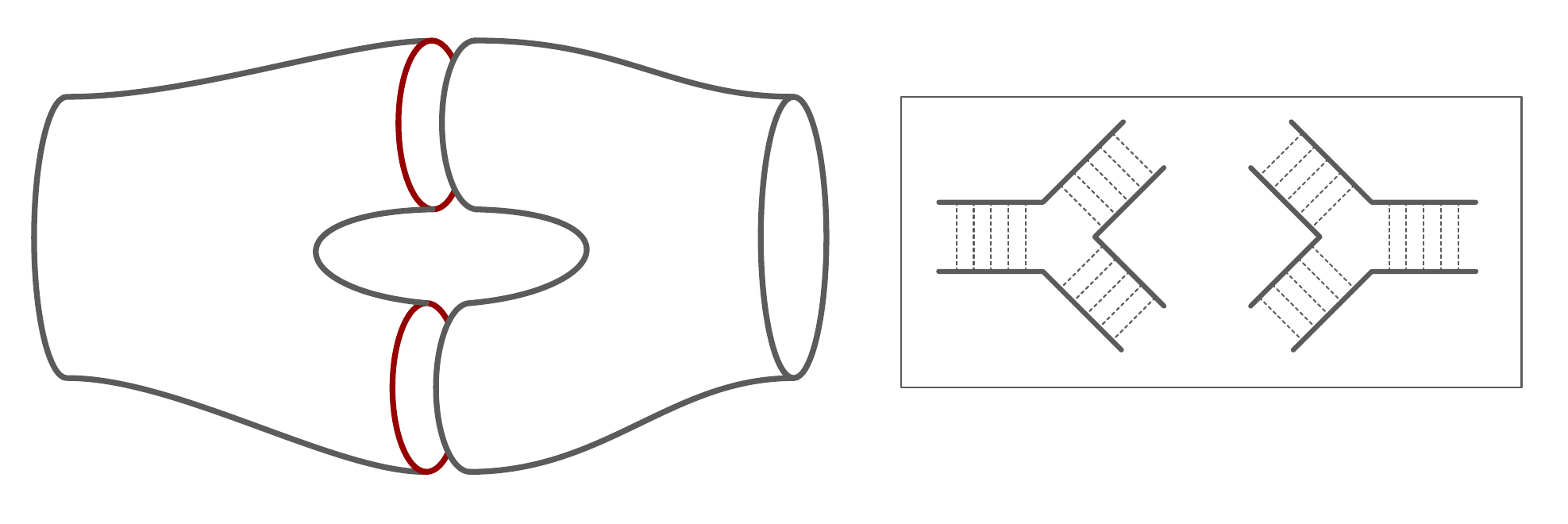}
\caption{The splitting off, and subsequent absorption, of a baby universe in JT
gravity. Inset: analogous process represented in matrix theory via the pairing
of two cubic vertices of the Kontsevich model.}
\label{fig:BabyUniverse}
\end{figure}

\subsubsection{Spectral curve and Calabi-Yau manifold}

Let us recall how the spectral density can be conveniently defined via the
associated spectral curve, $y(z)$, as described around Eqs.
\eqref{eq:SpectralCurveDefinition} and
\eqref{eq:SpectralDensityFromSpectralCurve}. In this section the spectral curve
will take on an important geometric meaning, in the construction of the UFT.
The approach may at first seem somewhat technical, but its ultimate power to
describe the non-perturbative bulk will be ample reward. We would like to
describe a theory capable of creating and destroying an arbitrary number of 2D
universes whose spatial topology is a circle. The multi-boundary correlations
are then multi-universe correlations, and topology change can be seen as one
universe splitting in two by emitting and later absorbing a `baby
universe'\footnote{This picture also explains the name Universe Field Theory.},
see Fig.~\ref{fig:BabyUniverse}. The key technical idea is that a two-dimensional JT universe
can by realised as the world-sheet of a judiciously chosen string theory, and
the process of baby universe creation and absorption is then given by string
amplitude techniques. We can construct the required string theory by specifying
the target space, which is a non-compact \keyword{Calabi-Yau (CY) manifold} defined by the
equation
\begin{equation}\label{eq:CYTargetSpace}
    H(x,y) - uv = 0\,,
\end{equation}
where $(x,y,u,v)\in \mathbb{C}^4$. Considered as one complex equation for four
complex variables, Eq.~\eqref{eq:CYTargetSpace} defines a manifold of real
dimension six. Our specific CY is defined in terms of a function  $H$  such that
\begin{equation}
    \Sigma: \qquad H(x,y) = 0
\end{equation}
is the spectral curve of JT gravity, given by $H(x,y) = y^2 - \frac{1}{16\pi^2}
\sin^2 \left( 2\pi \sqrt{2\phi_r x} \right)$. The coordinate $x$ here is a complex
variable defining the target space, but we can translate it into physical energy
of the spectrum via $x=-E$, giving rise to the JT energy density $\rho_0 =
\frac{1}{4\pi^2}\sinh\left( 2\pi\sqrt{2\phi_r E}\right),$ consistent with
Eq.~\eqref{eq:SpectralDensityFromSpectralCurve}. The physical spectral density Eq.~\eqref{eq:SpectralDensityJT} is
further multiplied by a factor of $e^{S_0}$. Near the spectral edge, that is for $E\rightarrow 0$, this has the square root
form $\rho_0 \sim \sqrt{E}$, and the corresponding `Airy' spectral curve $H(x,y)
= y^2 - x$. This limit has already been discussed in Section \ref{sec:spectralEdge}, in conjunction with the Kontsevich model. The complex manifold \eqref{eq:CYTargetSpace} is geometrically a
fibration over $\Sigma$ with the fibres degenerating on $\Sigma$. It gives rise
to the extra degrees of freedom needed to define the theory 
in the form of compact `color branes' and non-compact `flavor branes'. For our
current purposes, we do not have to deal with the higher-dimensional
CY, it suffices to integrate out its extra degrees of freedom
  reducing the theory
into an effectively two-dimensional description on $\Sigma$, in terms of a
two-dimensional version of \keyword{Kodaira-Spencer (KS) theory}. Technically,
this reduction assumes the form of  a chiral boson $\Phi(x)$ on
$\Sigma$ with propagator $\langle \Phi(x) \Phi(x') \rangle_{\rm KS} =
-\lambda_{\rm KS} \ln (x-x')$. Its action contains a cubic
vertex  with coupling constant $\lambda_{\rm KS} =
e^{-S_0}$ as a leading nonlinearity. Geometrically, this coupling describes the
mini-universe branching in Fig.~\ref{fig:BabyUniverse}; perturbation theory in
it quantitatively reproduces the topological expansion of JT gravity.~\cite{PostvanDerHeijdenVerlinde2022UniverseFieldTheory}

\subsubsection*{JT universes with and without boundaries}
The above construction  summarizes the
construction of the exact JT theory of 2D gravity on a string world-sheet, more
precisely a topological string world sheet, whose target space is the
CY-manifold \eqref{eq:CYTargetSpace}.
While the idea to identify JT universes with string world sheets may at first sound
strange,  this point of view is justified by its usefulness. 
Topological string theory is a
simplified version of string theory that ignores most physical details and
focuses only on the underlying geometric structure, especially the topology of
spacetime. A consistent anomaly-free theory of this type can be formulated only
if the geometry involved is Calabi-Yau manifold, justifying the appearance of
these objects in the present discussion. This theory has closed as well as open
string excitations, which in our interpretation correspond to JT universes with
suitable boundary terms added. Open strings (2D JT `universes') end on
structures known as D-branes, and KS theory offers two types,
\begin{enumerate}
    \item A large number $D=e^{S_0}$ of compact branes: These wrap compact
    $\mathbb{P}^1$ subspaces, `cycles', of the CY. The open JT-strings ending on
    them carry Chan-Paton color factors that are described by a $D\times D$
    color matrix $H_{ij}$. Integrating over these degrees of freedom gives rise
    to the SSS matrix model average, as seen in gravitational language. 
    \item An number $n \sim {\cal O}(1)$ of non-compact `flavor' branes: these
    wrap either the non-compact subspace $u=0, v\neq 0$, or $u\neq 0, v=0$ and
    the JT strings ending on them are given by insertions of det$(z-H)$ or
    det$(z-H)^{-1}$ in the matrix model generated by the non-compact branes. In
    other words, the non-compact branes add ratios of determinants that allow
    one to probe spectral correlations of the KS theory analogously to the
    $\sigma-$model treatment above.
\end{enumerate}
Note how the structure above suggests a new perspective of two-dimensional
     gravity as a geometric theory  of chaotic correlations. The construction of
     the KS theory on $\Sigma$ involves the integration over  an `ensemble' of
     open strings, coupling to the theory via the color matrix $\{H_{ij}\}$.  In this way, the emergence of
     gravity as an effectively ensemble averaged theory becomes 
     compelling. 

In addition to the open JT-string sector, there are also closed JT-strings, that
     is JT-unviverses without boundary whose spatial cross-section is an $S^1$. These are
     described by the chiral boson vertex operator $\partial \Phi(x)$ and they
     can split and join, governed by the closed-string coupling $\lambda_{\rm
     KS}$. In terms of the chiral boson langauge the non-compact brane
     determinants are represented by
\begin{equation}
    \psi(x) = e^{\Phi(x)} \leftrightarrow {\rm det}(x-H)\,,\qquad \psi^\dagger(z) = e^{-\Phi(z)} \leftrightarrow 1/{\rm det}(z-H)
\end{equation}
For example, we can write the resolvent in KS language as
\begin{equation}
    R(z) = \left\langle {\rm Tr} \frac{1}{z-H} \right \rangle = \left\langle \partial_z \Phi(z) \right\rangle_{\rm KS}\,,
\end{equation}
while multi-point spectral correlators are multi-brane correlators. Of special
interest are the determininat ratios 
\begin{equation}
    D_n = \left\langle \psi(x_1) \psi^\dagger(z_1) \cdots \psi(x_n) \psi^\dagger(z_n) \right\rangle_{\mathrm{KS}}\,,
\end{equation}
since they act as general functionals of $n-$point correlators of the spectral
density.

\begin{BoxTypeA}[box:annealed-quenched]{Annealed average? Or quenched?}

    Above we reasoned that the integral over the color matrix degrees of
    freedom, $\{H_{ij}\}$ introduces an effective ensemble average. Critical
    readers may object that the $H_{ij}$-amplitudes represent internal degrees of freedom
    of the theory. The `average' then is more  akin to an integral over
    environmental degrees of freedom, i.e. an \textit{annealed average} in the parlance of statistical
    physics. At the same time, we have seen that our observables of interest
    assume the form of brane/determinant ratios, structurally identical to
    the ratios in Eq.~\eqref{eq:SigmaDeterminantRatio}. The consequence of this
    identity is that the integral $\int DH \frac{\det(\dots)}{\det(\dots)}$ ---
    conceptually annealed --- 
    is mathematically equivalent to a quenched (ensemble) average in matrix
    theory. At this point, we do not understand if there is a deeper physicaly meaning to this equivalence, or if it is a mere mathematical coincidence. 
\end{BoxTypeA}

The key development of
\cite{AltlandPostSonnerVanDerHeijdenVerlinde2023QuantumChaos2DGravity} is to
show that within KS theory  the normal-ordered version of this correlator is
equivalent to the flavor-sigma model of Kontsevich type
\cite{AltlandPostSonnerVanDerHeijdenVerlinde2023QuantumChaos2DGravity}. More
specifically,  this reference expresses the  determinant correlator as  an
$(n|n)$ graded flavor matrix integral
\begin{equation}\label{eq:KSKontsevich}
\boxed{\Big\langle \big\{e^{\Phi(x_1)} e^{-\Phi(z_1)}  
\cdots e^{\Phi(x_n)} e^{-\Phi(z_n)}\big\}\Big\rangle_\mathsf{KS} =  
\int\limits_{(n|n)}^{\vphantom{|}} dA \,\exp\left[- e^{S_0}\left(\Gamma(A) - \mathrm{str}(XA)\right)\right]}
\end{equation}
with external sources $X = {\rm diag}(x_1\,,\ldots x_n | z_1\,,\ldots z_n)$. The
matrix potential $\Gamma(A)$ can be determined to any order in $\lambda_{\rm
KS}$ (see
\cite{AltlandPostSonnerVanDerHeijdenVerlinde2023QuantumChaos2DGravity}), but its
precise form is not needed here; all that matters is that  it starts
with a contribution of cubic order $\Gamma(A)\sim A^3+\dots$.

\subsection{Results}
\label{sec:StringTheoryResults}

The result Eq.~\eqref{eq:KSKontsevich}  establishes contact with our previous discussion of
near-edge spectral correlations in terms of the Kontsevich model and its
descendant $\sigma$-model, cf. Box~\ref{box:sigma-model}.  To repeat the main
conclusions in a language adapted to the present context,
\begin{itemize}
    \item  the Kontsevich
model~\eqref{eq:KSKontsevich} predicts a bulk microstate spectrum of almost
crystalline rigidity, as symptomatic for dense quantum systems. Mathematically,
this reflects in  the Airy formula~\eqref{eq:SpectralDensityEdgeDense}
with the identification $\Delta_0\sim \exp(-S_0 2/3)$ for the microstate
spacing. 
\item For energy parameters $X\gtrsim \Delta_0$, a stationary phase analysis
collapses the Kontsevich model to the nonlinear $\sigma$-model, which then
describes graviational spectral correlations in a regime equivalent to that of
the double scaled limit of matrix theory. 
\item In passing, we note that the Kontsevich model may also be applied to
describe the random geometry of quantum \textit{states} in the bulk Hilbert space. For a
discussion of these structures, we refer to Ref.~\cite{Altland:2025SpectralEdge}.
\end{itemize}

Let us conclude  this section with a final remark on dualities
between matrix theory and bulk geometric structures. In this review, we have
seen two types of  matrix ensembles  at work:  `color' matrix ensembles of high
dimensions $\sim\exp(S_0)$ as proxies of the geometric structures otherwise
described by the JT or KS functional. And `flavor' (super-)matrix ensembles of
low dimension $\sim n$, representing the number of spectral probe insertions into
the theory. In the theory of quantum chaos, these two descriptions of chaotic
correlations are related to each other via the celebrated color-flavor duality,
technically introduced in terms of  Hubbard-Stratonovich transformations. 

In a miraculous way, an equivalent color-flavor duality emerges in the
dimensional reduction of the CY-manifold to the spectral curve $\Sigma$. Here, the
two branches of the theory emerge upon wrapping the six-dimensional CY-manifold
either in terms of color- or flavor-branes as outlined above. Concerning the
flavor reduction, elements of the supersymmetric $\sigma$-model which in the
historical construction~\cite{Efetbook} appeared on the basis of algebraic principles (the
flavor matrix manifold with its different saddle points, the combination of
compact and non-compact coordinate directions, etc.) now all acquire a 
geometric interpretation in terms of stringy degrees of freedom. At this point,
we know how to describe this reduction technically, but cannot
say to fully comprehend its physical meaning. In particular, it will be
interesting to explore if similar `geometric seeds of quantum chaos' are
hardwired into higher dimensional variants of string theory. 

\section{Discussion}
\label{sec:Discussion}
We started this review with an introduction to the SYK model as a proposed
boundary dual of two-dimensional gravity. This connection was corroborated via
several lines of evidence, matching spectral densities, the Schwarzian action as
a common low energy effective action, and random matrix correlations in the
respective spectra. At the same time, it has been pointed out early on that  the
relation between SYK and JT gravity is an IR one: JT captures the Schwarzian
soft sector of SYK, but not its full microscopic UV
completion.~\cite{Blommaert:2024DilatonHologram,Maldacena:2016upp} In particular,
beyond the Schwarzian regime one expects additional structure, and in solvable
limits such as double-scaled SYK the relevant bulk completion is not pure JT
gravity. While one might dismiss these differences as `UV-effects', our
subsequent discussion  of the gravitational path integral revealed an independent
`IR-issue', which we believe affects the theory at its core: the SYK model is a
sparse chaotic system, polynomially many (in $N$)  parameters seeding randomness
into an exponentially high dimensional Hilbert space. As observable consequences
we discussed strong collective fluctuations of the spectral density,
specifically at the spectral edge. By contrast, JT gravity, and its string
theoretic extension are dense, the number of random parameters in their dual
matrix ensembles even exceeds the Hilbert space dimension. As a consequence, the
edge of the gravitational spectrum is defined with level spacing
precision.~\cite{Altland:2025SpectralEdge} This is a disturbing observation  as it
implies that no `conventional' many-body system, i.e. one governed by
interactions of low operator contents and therefore sparse by construction will
qualify as a microscopic boundary partner. 

While the story of the duality between JT gravity, or its non-perturbative
completion in terms of KS theory, and random matrix theory is rather complete,
some open questions remain that deserve further attention, even if we keep the
focus of our attention on the 2D story. The first is technical: can the KS
completion of JT be shown to be in some sense unique or does non-perturbative
ambiguity remain? Relatedly, the chaos angle on random matrix theory initially
motivated the search for a {\it single} unitary quantum chaotic (boundary)
Hamiltonian, whose spectral correlations reproduce the JT matrix model when
averaged in a suitable manner, e.g. over energy windows, see for example
~\cite{Haneder:2024SchwarzianDensity}. Note, however, that our discussion of the collective fluctuations of the edge
spectrum in sparse quantum systems above gives strong constraints on any
possible such Hamiltonian, cutting down significantly the phase space of
potential models. Alternatively, from the bulk perspective
one would like to know which modifications and additions to JT gravity would
cause the bulk theory to
factorise~\cite{Blommaert:2019wfy,Mukhametzhanov:2021nea,Blommaert:2021fob,Blommaert:2022ucs,Saad:2021rcu,Boruch:2024kvv}.
However,  attempts to modify the bulk often have exotic
side-effects such as severe non-locality. From a bulk gravity perspective, the
goal is to find a modification that is sufficiently mild so that the resulting
theory still preserves the essential features of `pure 2D gravity', but gives
factorized answers for quantities, such as ${\cal Z}(\beta_1){\cal
Z}(\beta_2)\Bigr|_{\rm bulk}$. In our non-perturbative language the
corresponding question is what would be the minimal ingredients that should be
added to JT / KS theory, so as to `factorize' the ensemble, in the sense that
connected correlations of the form \eqref{eq:connectedPartitionFunction} vanish,
and the necessity of a boundary ensemble disappears.

At this point,  the perhaps most important and ambitious endeavour
is to extend the new insights and techniques that the quantum chaos perspective
has afforded to higher dimensions, starting with 3D. A promising avenue is
pursued in~\cite{Jafferis:2022uhu}, which ports the constraint
matrix model approach to the SSS model, \cite{Jafferis:2022JTMatter} to the
arena of AdS$_3/$CFT$_2$. The idea is to construct a random matrix-like
structure that realizes the maximum ignorance ensemble on the data that defines
a boundary CFT, subject only to the constraints of modularity and crossing of
correlation functions. This gives rise to a quartic matrix/tensor model which
has shown rich promise as a theory of pure 3D gravity based on triangulating 3D
spacetime~\cite{Cotler:2018zff,Cotler:2020ugk,Belin:2020hea,Belin:2020jxr,Belin:2023efa,Jafferis:2022uhu,Collier:2023fwi,Collier:2024mgv}.
Concerning the boundary, the extension of single SYK cells to arrays
with interactions engineered to introduce a sense  of
chirality~\cite{AltlandCallebaut25} is governed by a generalization of the
Schwarzian action to so-called Alekseev-Shatashvili theory~\cite{Alekseev:1988ce}. The same theory
features as the boundary action of the Chern-Simons theory describing 3D black
hole solutions~\cite{Cotler:2018zff}, establishing a link similar to that between JT gravity and the
Schwarzian.

More generally, 
the quantum chaos approach to quantum gravity challenges us to define and
understand the notion of chaotic conformal field theory.
~\cite{Perlmutter:2016pkf,Haehl:2018izb,Haehl:2023tkr,Belin:2020hea,Belin:2023efa}.
This exciting new research direction 
suggests deep mathematical structures needed to reconcile the strong
constraints inherent in CFTs, especially in two dimensions, with the equally
strong constraints of hard quantum chaos. Despite the richness of the structure
already discovered, especially in low dimensions, it feels that so far we have
only touched the tip of the iceberg of the beautiful and fascinating interplay
of quantum chaos and quantum gravity.

\section*{Acknowledgements}
We would like to thank our collaborators over the years, on these and related topics, for their invaluable contributions to the material presented in this review.
The work in this review was in part supported by
the Fonds National Suisse de la Recherche Scientifique through Project Grant 200021\_215300, the NCCR ``The Mathematics of Physics (SwissMAP)'' (NCCR 51NF40-141869), as well as the Swiss State Secretariat for Education, Research and Innovation (SERI) under contract number UeMO19-5.1.

\bibliographystyle{JHEP}
\bibliography{HolographyChaosReview}

\end{document}